\documentclass[aps,prc,twoside,twocolumn,nofootinbib,10pt,showpacs,floatfix]{revtex4-1}
\usepackage{amsmath,amssymb}
\usepackage{graphicx,bm}
\usepackage{slashed}
\usepackage{epstopdf}
\usepackage{ulem} 
\usepackage[usenames]{color}
\usepackage{float}
\usepackage{hyperref}
\usepackage{subfigure}
\usepackage{subfigure}
\usepackage{rotating}
\usepackage{color}
\usepackage{multirow}
\usepackage{dcolumn}
\usepackage{overpic}
\usepackage{booktabs}
\usepackage{makecell}
\usepackage{color}
\usepackage{diagbox}

\renewcommand\sout{\bgroup \color{red} \ULdepth=-.5ex \ULset}

\begin{document}
\title{Fully heavy pentaquarks}
\author{Hong-Tao An$^{1,5}$}\email{anht14@lzu.edu.cn}
\author{Kan Chen$^{3,4,1,5}$}\email{chenk$_$10@pku.edu.cn}
\author{Zhan-Wei Liu$^{1,2,5}$}\email{liuzhanwei@lzu.edu.cn}
\author{Xiang Liu$^{1,2,5}$}\email{xiangliu@lzu.edu.cn}
\affiliation{
$^1$School of Physical Science and Technology, Lanzhou University, Lanzhou 730000, China\\
$^2$Lanzhou Center for Theoretical Physics, Key Laboratory of Theoretical Physics of Gansu Province, and Frontiers Science Center for Rare Isotopes, Lanzhou University, Lanzhou 730000, China\\
$^3$Center of High Energy Physics, Peking University, Beijing 100871, China\\
$^4$School of Physics and State Key Laboratory of Nuclear Physics and Technology, Peking University, Beijing 100871, China\\
$^5$Research Center for Hadron and CSR Physics, Lanzhou University and Institute of Modern Physics of CAS, Lanzhou 730000, China}

\date{\today}
\begin{abstract}
Very recently, the LHCb Collaboration reported a fully charmed tetraquark state $X(6900)$ in the invariant mass spectrum of $J/\psi$ pairs. If one $J/\psi$ meson is replaced with a fully charmed baryon, we obtain a fully charmed pentaquark candidate. In this work, we perform a systematic study on the mass spectra of the S-wave fully heavy pentaquark $QQQQ\bar{Q}$ in the framework of the chromomagnetic interaction model. Based on our results in two different schemes, we further investigate the decay behaviors for them. We hope that our study will be helpful in searching for such types of exotic pentaquark states in experiments in the future.
\end{abstract}
\maketitle

\section{Introduction}\label{sec1}

At the birth of the quark model \cite{GellMann:1964nj,Zweig:1981pd,Zweig:1964jf}, Gell-Mann and Zweig indicated that hadronic states with the $qq\bar{q}\bar{q}$ and $qqqq\bar{q}$ quark configurations should exist in nature. Such exotic states were further investigated with some phenomenological models soon afterwards. For example, Jaffe adopted the quark-bag model to study $q^2\bar{q}^2$ hadrons, where the mass spectrum and dominant decay behavior were predicted \cite{Jaffe:1976ig}. In 1979, Strottman calculated the masses of $q^4\bar{q}$ and $q^5 \bar{q}^2$ in the framework of the MIT bag model \cite{Strottman:1979qu}.
{The name {\it pentaquark} was firstly proposed by Gignoux et al. in 1987 \cite{Gignoux:1987cn,Lipkin:1987sk},
and they found that the states $P^{0}=\bar{c}uuds$ and $P^{-}=\bar{c}ddus$ with spin 1/2 and their beauty analogs are very likely to be stable pentaquarks.}
Since 2003, with the accumulation of experimental data, more and more charmonium-like $XYZ$ states were reported in experiments. Especially, the observation by the LHCb Collaboration confirms the existence of pentaquark $P_c$ states \cite{Aaij:2015tga,Aaij:2016phn,Aaij:2019vzc}. In the past 
 twenty years, progress has been made on studying exotic multiquarks \cite{Chen:2016qju,Liu:2019zoy,Guo:2017jvc,Brambilla:2019esw}. 

Recently, the LHCb Collaboration studied the invariant mass spectrum of $J/\psi$ pairs, and they reported a narrow structure around 6.9 GeV and a broad structure in the mass range 6.2-6.8 GeV. The global significance for these two structures are larger than 5$\sigma$. Such distinct structures are expected to be with the $cc\bar{c}\bar{c}$ configuration \cite{Aaij:2020fnh}.

The $cc\bar{c}\bar{c}$ tetraquark had been discussed a lot in the literature before its discovery by the LHCb Collaboration. In Ref. \cite{Iwasaki:1975pv}, a $cc\bar{c}\bar{c}$ tetraquark was predicted at about 6.2 GeV.  The $cc\bar{c}\bar{c}$ tetraquark system was systematically investigated based on a quark-gluon model in Ref. \cite{Chao:1980dv}.
{
In Ref. \cite{Ader:1981db} Ader et al. found the lowest $cc\bar{c}\bar{c}$ state is not bound, 
and there are other discussions with a similar conclusion \cite{,Chiu:2005ey,Hughes:2017xie,SilvestreBrac:1992mv,Vega-Morales:2017pmm,Chen:2016jxd,Richard:2017vry}.}


This important signal from the LHCb Collaboration provides us a new ground to understand the nonperturbative behavior of QCD \cite{Chao:2020dml,Richard:2020hdw,Maiani:2020pur}. The mass of the observed signal around 6.9 GeV is consistent with a previous QCD sum rule predictions \cite{Chen:2016jxd}. After the LHCb Collaboration reported their results, the strong decay properties of $S$- and $P$-wave tetraquark states were further studied in Ref. \cite{Chen:2020xwe}, and the observed structure at around 6.9 GeV is suggested to be with $J^{PC}=0^{-+}$ or $1^{-+}$. Becchi et al. have studied that the tetraquarks $cc\bar{c}\bar{c}$ with $J^{PC}=0^{++}$, $2^{++}$ decay into four mouns and into hidden- and open-charmed mesons and provide the decay widths of fully charmed tetraquarks \cite{Becchi:2020uvq}. The inner structures of the fully charmed tetraquark state were studied \cite{Guo:2020pvt}. The mass spectrum of $cc\bar{c}\bar{c}$ tetraquarks were studied in an extended relativized quark model, QCD sum rule, chromomagnetic model and so on \cite{Lu:2020cns,Albuquerque:2020hio,Wang:2020dlo,Zhang:2020xtb,Giron:2020wpx,Faustov:2020qfm,Gordillo:2020sgc,Weng:2020jao}. The production mechanism of $cc\bar{c}\bar{c}$ was also studied in various schemes \cite{Wang:2020gmd,Wang:2020wrp,Dong:2020nwy,Maciula:2020wri,Karliner:2020dta,Feng:2020riv,Ma:2020kwb,Zhu:2020xni,Zhu:2020xni}.

The discoveries of fully heavy tetraquark states and $P_c$ states make one speculate that the pentaquark state with fully heavy quarks $QQQQ\bar{Q}$ may also exist. If the mass is above the  baryon-meson thresholds, the heavy pentaquark state may allow the strong decays into the corresponding two body. The study of the masses and decay properties would help to search for the heavy pentaquark states in experiments.

The strong interaction in a fully heavy multiquark state is not clear at present. The chromomagnetic interaction (CMI) model provides us with a simple picture to quantitatively understand the spectrum of multiquark states. In the framework of the CMI model \cite{DeRujula:1975qlm}, the strong interaction between quarks via gluon exchange force is parametrized into effective quark masses and quark coupling parameters. Despite its simple Hamiltonian, this model can catch the basic features of hadron spectra, since the mass splittings between hadrons reflect the basic symmetries of their inner structures \cite{SilvestreBrac:1992mv}. This model has been widely adopted to study the mass spectra of multiquark systems \cite{
Wu:2016vtq, Wu:2016gas, Chen:2016ont, Wu:2017weo, Luo:2017eub, Zhou:2018pcv, Li:2018vhp,Wu:2018xdi, An:2019idk,Weng:2019ynva,Cheng:2019obk,Liu:2016ogz,Hogaasen:2013nca,Cheng:2020irt,Cheng:2020nho,Cheng:2020wxa,Weng:2018mmf,Weng:2020jao,Zhao:2014qva,Wu:2017weo,Zhou:2018pcv,Li:2018vhp,An:2019idk}.
In this work, we systematically study the S-wave $QQQQ\bar{Q}$ pentaquark system within the framework of the CMI model to calculate the mass spectra and the relative partial decay widths,  and find a possible stable pentaquark state.

This paper is organized as follows.
In Sec. \ref{sec2}, we introduce the CMI model and determine the relevant parameters used in the CMI model.
The $flavor \otimes color \otimes spin$ wave functions are constructed and the CMI Hamiltonian elements are calculated for the $QQQQ\bar{Q}$ pentaquark system in Sec. \ref{sec3}.
In Sec. \ref{sec7}, we present the mass spectra, the mass splittings, the possible strong decay channels, and the relative partial decay widths, and also discuss the stability for the pentaquark states.
A short summary is followed in Sec. \ref{sec8}. 
Finally, some useful expressions are presented in the Appendix.


\section{The Chromomagnetic Interaction Model}\label{sec2}
The masses of the ground hadrons can be obtained by the effective Hamiltonian at quark level
\begin{eqnarray}\label{Eq1}
H&=&\sum_im_i+H_{\textrm{CMI}} \nonumber \\
&=&\sum_im_i-\sum_{i<j}C_{ij} \vec\lambda_i\cdot \vec\lambda_j \vec\sigma_i\cdot\vec\sigma_j,
\end{eqnarray}
where the $H_{\textrm{CMI}}$ denotes the Hamiltonian of the chromomagnetic interaction \cite{DeRujula:1975qlm}.
$\sigma_{i}$ and $\lambda_i$ are the Pauli matrices and the Gell-Mann matrices, respectively.
For the antiquark, the $\lambda_i$ should be replaced with $-\lambda_i^{*}$.
$m_i$ is the effective mass of the $i$-th constituent quark.
In the above Hamiltonian, the chromoelectric interaction and color confinement effect are also incorporated in the effective quark mass $m_i$.
$C_{ij}$ is the effective coupling constant between the $i$-th quark and $j$-th quark.
The effective quark mass $m_i$ and the coupling constant $C_{ij}$ can be determined from the experimental hadron masses.

As indicated in Refs. \cite{Wu:2016vtq, Wu:2016gas, Chen:2016ont, Wu:2017weo, Luo:2017eub, Zhou:2018pcv, Li:2018vhp, Wu:2018xdi,An:2019idk,Liu:2016ogz},
the predicted hadron masses obtained from Eq. (\ref{Eq1}) are generally overestimated.
The main reason is that the dynamical effects inside hadrons can not simply be absorbed into the effective quark masses.
Thus, in order to take such effective interaction into account, we replace the sum of the $m_i$ term in Eq. (\ref{Eq1}) with $M_{ref}-\langle H_{\rm CMI}\rangle_{ref}$, where $M_{ref}$ is a reference mass scale and $\langle H_{\rm CMI}\rangle$ is the corresponding CMI matrix element.
The mass of the ground hadron can thus be written as
\begin{eqnarray}
\label{Eq11}
M=M_{ref}-\langle H_{\rm CMI}\rangle_{ref}+\langle H_{\rm CMI}\rangle.
\end{eqnarray}
Here, we choose the baryon-meson thresholds as the mass scales, where
the reference baryon-meson system should have the same constituent quarks with the studied pentaquark state.
In this way, the dynamical effects that are not incorporated in the original approach are somehow phenomenologically compensated for this procedure \cite{Zhou:2018pcv}.
We label this method as the reference mass scheme.

In addition, we introduce another scheme to estimate the masses of pentaquark states.
We separate the two-body chromoelectric effects out of the effective quark masses and generalize the chromomagnetic interaction model by writing the chromoelectric term explicitly \cite{Hogaasen:2013nca,Weng:2019ynva,Weng:2018mmf,Weng:2020jao} i.e.,
\begin{eqnarray}\label{Eq3}
H&=&\left(\sum_im_i^0+H_{\textrm{CEI}}^0\right)+H_{\textrm{CMI}}^0\nonumber\\
&=&\sum_im_i^0-\sum_{i<j}A_{ij} \vec\lambda_i\cdot \vec\lambda_j-\sum_{i<j}v_{ij} \vec\lambda_i\cdot \vec\lambda_j \vec\sigma_i\cdot\vec\sigma_j, \nonumber\\
&=&-\frac{3}{4}\sum_{i<j}m_{ij}\vec\lambda_i\cdot \vec\lambda_j-\sum_{i<j}v_{ij}\vec\lambda_i\cdot \vec\lambda_j \vec\sigma_i\cdot\vec\sigma_j+...
\end{eqnarray}
where the omitted operator nullifies the color-singlet physical states, and
\begin{eqnarray}\label{Eq4}
m_{ij}=\frac{1}{4}(m_{i}^0+m_{j}^0)+\frac{4}{3}A_{ij}.
\end{eqnarray}

This treatment has been successfully adopted in Refs. \cite{Karliner:2014gca,Karliner:2016zzc,Karliner:2017qjm,Hogaasen:2013nca,Weng:2019ynva,Weng:2018mmf,Weng:2020jao}.
The parameters $m_{ij}$ and $v_{ij}$ are also determined from the experimental hadron masses.
In this work, we label this method as the modified CMI model scheme.

To estimate the masses of the $QQQQ\bar{Q}$ pentaquark states,
we need some hadron masses as input to fit the effective coupling parameters $C_{ij}$, $m_{ij}$, and $v_{ij}$  \cite{Tanabashi:2018oca}.
These conventional hadrons are listed in Table \ref{comp}.
Because some of the heavy flavor baryons are not yet observed, we introduce the theoretical results in Refs. \cite{Weng:2018mmf,Godfrey:1985xj} as our input, and enclose the theoretical values of masses for these baryons with parentheses in Table \ref{comp}.


\begin{table}[htbp]
\caption{The masses of conventional hadrons used for determining parameters in units of MeV \cite{Tanabashi:2018oca}.
The masses of not-yet-observed baryons and $B_{c}^{*}$ in parentheses are taken from Refs. \cite{Weng:2018mmf} and \cite{Godfrey:1985xj}, and others are from experiments.}\label{comp}
\begin{tabular}{crc|crc}
\bottomrule[1.5pt]
\bottomrule[0.5pt]
hadrons& $I(J^P)$ &Mass & hadrons& $I(J^P)$ &Mass \\
\bottomrule[0.7pt]
$\eta_{c}$&$0(0^{-})$&2983.9&$\Omega_{ccc}$&$0(3/2^{+})$&(4785.6)\\
$J/\psi$&$0(1^{-})$&3096.9&$\Omega_{ccb}$&$0(1/2^{+})$&(7990.3)\\
$\eta_{b}$&$0(0^{-})$&9399.0&$\Omega_{ccb}^{*}$&$0(3/2^{+})$&(8021.8)\\
$\Upsilon$&$0(1^{-})$&9460.3&$\Omega_{bbc}$&$0(1/2^{+})$&(11165.0)\\
$B_{c}$&$0(0^{-})$&6274.9&$\Omega_{bbc}^{*}$&$0(3/2^{+})$&(11196.4)\\
$B_{c}^{*}$&$0(1^{-})$&(6338.0)&$\Omega_{bbb}$&$0(3/2^{+})$&(14309.7)\\
$\Xi_{cc}$&$1/2(1/2^{+})$&3621.4&$\Xi_{bb}$&$1/2(1/2^{+})$&(10168.9)\\
$\Xi^{*}_{cc}$&$1/2(3/2^{+})$&(3696.1)&$\Xi^{*}_{bb}$&$1/2(3/2^{+})$&(10188.8)\\
$\Omega_{cc}$&$0(1/2^{+})$&(3731.8)&$\Omega_{bb}$&$0(1/2^{+})$&(10259.0)\\
$\Omega^{*}_{cc}$&$0(3/2^{+})$&(3802.4)&$\Omega^{*}_{bb}$&$0(3/2^{+})$&(10267.5)\\
$\Xi_{cb}$&$1/2(1/2^{+})$&(6922.3)&$\Omega_{cb}$&$0(1/2^{+})$&(7010.7)\\
$\Xi'_{cb}$&$1/2(1/2^{+})$&(6947.9)&$\Omega'_{cb}$&$0(1/2^{+})$&(7047.0)\\
$\Xi^{*}_{cb}$&$1/2(3/2^{+})$&(6973.2)&$\Omega^{*}_{cb}$&$0(3/2^{+})$&(7065.7)\\
\bottomrule[0.5pt]
\midrule[1.5pt]
\end{tabular}
\end{table}


\begin{table}[t]
\centering \caption{Coupling parameters for the schemes in units of MeV.
}\label{parameter2}
\renewcommand\arraystretch{1.25}
\begin{tabular}{cccccc}
\bottomrule[1.5pt]
\bottomrule[0.5pt]
\multicolumn{6}{l}{The reference mass scheme:}\\
$C_{cc}$&$C_{bb}$&$C_{cb}$&$C_{c\bar{c}}$&$C_{b\bar{b}}$&$C_{c\bar{b}}$\\
3.3&1.8&2.0&5.3&2.9&3.3\\
\bottomrule[1.0pt]
\multicolumn{6}{l}{The modified CMI model scheme:}\\
$m_{cc}$&$m_{bb}$&$m_{cb}$&$m_{c\bar{c}}$&$m_{b\bar{b}}$&$m_{c\bar{b}}$\\
792.9&2382.4&1604.0&767.1&2361.2&1580.6\\
\bottomrule[0.5pt]
$v_{cc}$&$v_{bb}$&$v_{cb}$&$v_{c\bar{c}}$&$v_{b\bar{b}}$&$v_{c\bar{b}}$\\
3.5&1.9&2.0&5.3&2.9&2.9\\
\bottomrule[0.5pt]
\bottomrule[1.5pt]
\end{tabular}
\end{table}

{
In principle, the values of $A_{ij}$ and $v_{ij}$ in the modified CMI model should be different for various systems. 
However, it is difficult to exactly calculate these parameters for a given system without knowing the spatial wave
function. Thus, they are extracted from the masses of conventional hadrons by assuming that quark-(anti)quark interactions are the same for all the hadron systems. Of course, this assumption certainly leads to uncertainties on mass estimations for multiquark states. Since the size of a multiquark state is expected to be larger than that of a conventional hadron and the distance between quark components may be larger, the attraction between quark components should be weaker. Thus, our framework may produce a little more binding.
}



Now we fit the effective coupling parameters $C_{ij}$, $m_{ij}$, and $v_{ij}$ in the reference mass and modified CMI model schemes by applying Eq. (\ref{Eq11}) and Eq. (\ref{Eq3}), respectively.
We present the obtained effective coupling parameters of $QQQQ\bar{Q}$ pentaquark states in Table \ref{parameter2}.
One can refer to Refs. \cite{Weng:2019ynva,Weng:2018mmf,Liu:2019zoy} for more details.

\section{The $QQQQ\bar{Q}$ pentaquark wave functions and The CMI Hamiltonian}\label{sec3}
In order to systematically study the mass spectra of the $QQQQ\bar{Q}$ pentaquark system, we need construct the wave function of $QQQQ\bar{Q}$ pentaquark first. We exhaust all the possible color $\otimes$ spin wave functions of pentaquark states, and combine them with the corresponding flavor wave functions. The constructed pentaquark wave functions should be constrained appropriately by the Pauli principle. After that, we can use these pentaquark wave functions to calculate the mass spectra of the corresponding pentaquark states.

The total wave function of the S-wave $QQQQ\bar{Q}$ pentaquark can be described by the direct product of flavor, color, and spin wave functions,
\begin{eqnarray}
\psi_{\textrm{tot}}=\psi_{flavor}\otimes\psi_{color}\otimes\psi_{spin}.
\end{eqnarray}
Due to the Pauli principle, this wave function should be fully antisymmetric when exchanging identical quarks.

In the flavor space, we divide the $QQQQ\bar{Q}$ pentaquark system into three groups of subsystems according to their symmetries: 
(1) the first four quarks are identical:  the $cccc\bar{c}$, $cccc\bar{b}$, $bbbb\bar{c}$, and $bbbb\bar{b}$ pentaquark subsystems,
(2) the first three quarks are identical: the $cccb\bar{c}$, $cccb\bar{b}$, $bbbc\bar{c}$, and $bbbc\bar{b}$ pentaquark subsystems,
(3) there are two pairs of identical quarks:  the $ccbb\bar{c}$ and $ccbb\bar{b}$ pentaquark subsystems.

In the color space, the color wave functions are singlets due to the color confinement.
The color wave functions can be deduced from the following direct product
\begin{eqnarray}\label{colorproduct}
\left[3_c\otimes3_c\otimes3_c\otimes3_c\right]\otimes\bar{3}_c
\end{eqnarray}
Based on Eq. (\ref{colorproduct}), the first four quarks are in three color triplet, the corresponding Young tableau [2,1,1] can be written as
\begin{align}\label{eq-color1}
\begin{tabular}{|c|c|}
\hline
1 &  2   \\
\cline{1-2}
\multicolumn{1}{|c|}{3} \\
\cline{1-1}
\multicolumn{1}{|c|}{4}  \\
\cline{1-1}
\end{tabular}_{\begin{tabular}{|c|} \multicolumn{1}{c}{$C_1$}\end{tabular}},
\quad
\begin{tabular}{|c|c|}
\hline
1 &  3    \\
\cline{1-2}
\multicolumn{1}{|c|}{2} \\
\cline{1-1}
\multicolumn{1}{|c|}{4}  \\
\cline{1-1}
\end{tabular}_{\begin{tabular}{|c|} \multicolumn{1}{c}{$C_2$}\end{tabular}},
\quad
\begin{tabular}{|c|c|}
\hline
1 &  4   \\
\cline{1-2}
\multicolumn{1}{|c|}{2} \\
\cline{1-1}
\multicolumn{1}{|c|}{3} \\
\cline{1-1}
\end{tabular}_{\begin{tabular}{|c|} \multicolumn{1}{c}{$C_3$}\end{tabular}}.
\end{align}
Then, by combining the antitriplet from the antiquark with the deduced three color triplets in Eq. (\ref{eq-color1}), we obtain
three color singlets for  the $QQQQ\bar{Q}$ pentquark system.


In the spin space, the spin state can be represented in terms of a five-dimensional Young tableau [3,2] as
\begin{align}
\label{eq-spin3}
&\begin{tabular}{|c|c|c|}
\hline
1&2&3  \\
\cline{1-3}
4&5 \\
\cline{1-2}
\end{tabular}_{\begin{tabular}{|c|} \multicolumn{1}{c}{$S_1$}\end{tabular}},
\begin{tabular}{|c|c|c|}
\hline
1&2&4  \\
\cline{1-3}
3&5 \\
\cline{1-2}
\end{tabular}_{\begin{tabular}{|c|} \multicolumn{1}{c}{$S_2$}\end{tabular}},
\begin{tabular}{|c|c|c|}
\hline
1&3&4  \\
\cline{1-3}
2&5 \\
\cline{1-2}
\end{tabular}_{\begin{tabular}{|c|} \multicolumn{1}{c}{$S_3$}\end{tabular}},
\begin{tabular}{|c|c|c|}
\hline
1&2&5  \\
\cline{1-3}
3&4 \\
\cline{1-2}
\end{tabular}_{\begin{tabular}{|c|} \multicolumn{1}{c}{$S_4$}\end{tabular}},
\begin{tabular}{|c|c|c|}
\hline
1&3&5  \\
\cline{1-3}
2&4 \\
\cline{1-2}
\end{tabular}_{\begin{tabular}{|c|} \multicolumn{1}{c}{$S_5$}\end{tabular}}.\\ \nonumber
\end{align}
for the pentaquark states with total spin $J=1/2$.
Since particle 5 is an antiquark, we can isolate this antiquark and discuss the symmetry property of the first four quarks: 1, 2, 3, and 4 in color $\otimes$ spin space.

When antiquark 5 is separated from the spin wave functions, 
the spin states represented in Young tableaus without antiquark 5 can be directly obtained from Eq. (\ref{eq-spin3}) as
\begin{align}
\label{eq-spin4}
&\begin{tabular}{|c|c|c|}
\hline
1&2&3  \\
\cline{1-3}
4 \\
\cline{1-1}
\end{tabular}_{\begin{tabular}{|c|} \multicolumn{1}{c}{$S_1$}\end{tabular}},
\begin{tabular}{|c|c|c|}
\hline
1&2&4  \\
\cline{1-3}
3 \\
\cline{1-1}
\end{tabular}_{\begin{tabular}{|c|} \multicolumn{1}{c}{$S_2$}\end{tabular}},
\begin{tabular}{|c|c|c|}
\hline
1&3&4  \\
\cline{1-3}
2 \\
\cline{1-1}
\end{tabular}_{\begin{tabular}{|c|} \multicolumn{1}{c}{$S_3$}\end{tabular}},
\begin{tabular}{|c|c|}
\hline
1&2  \\
\cline{1-2}
3&4 \\
\cline{1-2}
\end{tabular}_{\begin{tabular}{|c|} \multicolumn{1}{c}{$S_4$}\end{tabular}},
\begin{tabular}{|c|c|}
\hline
1&3 \\
\cline{1-2}
2&4 \\
\cline{1-2}
\end{tabular}_{\begin{tabular}{|c|} \multicolumn{1}{c}{$S_5$}\end{tabular}}.
\end{align}
We can identify the spin states in Eq. (\ref{eq-spin4}) with the Young-Yamanouchi bases for Young tableauss [3,1] and [2,2].
The pentaquark states with total spin $J=5/2,3/2$ have similar situations.

With the above preparation, we can start to construct the flavor $\otimes$ color $\otimes$ spin wave functions of $QQQQ\bar{Q}$ pentaquark states \cite{Stancu:1999qr,Park:2017jbn,Park:2015nha,Park:2016mez,Park:2016cmg,Park:2018oib}.
Then, based on the possible $\psi_{flavor}\otimes\psi_{color}\otimes\psi_{spin}$ bases of the $QQQQ\bar{Q}$ pentaquark system, we calculate the CMI matrices for the corresponding pentaquark states.
In Table \ref{nnnsQ} of the Appendix, we only present the expressions of CMI Hamiltonians for the $cccc\bar{c}$, $cccb\bar{c}$, and $ccbb\bar{c}$ pentaquark subsystems.
{Moreover, we also get more tractable CMI matrices under the $ccc \otimes b\bar{c}$ ($cc \otimes bb \otimes \bar{c}$) bases for the $cccb\bar{c}$ ($ccbb\bar{c}$) subsystem.
One can refer to Refs. \cite{Wu:2017weo,Zhou:2018pcv} for more details.}
As for the expressions of CMI matrices for the $cccc\bar{b}$, $bbbb\bar{c}$, $bbbb\bar{b}$, $cccb\bar{b}$, $bbbc\bar{c}$, $bbbc\bar{b}$, and $ccbb\bar{b}$ pentaquark subsystems,
we can obtain them from the expressions of the $cccc\bar{c}$, $cccb\bar{c}$, and $ccbb\bar{c}$ pentaquark subsystems according to their similar symmetry properties.

\section{Mass Spectra and Decay Behaviors}\label{sec7}
The interacting Hamiltonians can be diagonalized and one can thus obtain the eigenvalues as well as eigenvectors for the corresponding pentaquark systems. According to our results, we discuss the mass gaps, decay behaviors, and stabilities of all the $QQQQ\bar{Q}$ pentaquark states.

Based on the two schemes proposed in Sec. \ref{sec2}, we present the mass spectra for all the $QQQQ\bar{Q}$ pentaquark subsystems in Table \ref{mass-QQQQQ}. Take the $cccb\bar{c}$ pentaquark subsystem as an example. In the reference mass scheme, we use two types of baryon-meson reference systems ($\Omega_{ccc}+B_{c}$ and $\Omega_{ccb}+\eta_c$) to estimate the masses of $cccb\bar{c}$ states. Some results calculated from the two reference systems differ by more than a hundred MeV for the $cccb\bar{c}$ pentaquark states. However, the gaps with different reference systems are still the same. Thus, if one pentaquark state were observed, its partner states may be searched for with the relative positions presented
in Table \ref{mass-QQQQQ}. Such a study can be used to test our calculation. Here, we need to emphasis that as a rough estimation, the dynamics and contributions from other terms in the interacting potential are not elaborately considered in Eq. (\ref{Eq3})  \cite{Zhou:2018pcv}.

\begin{figure*}[t]
\begin{tabular}{cc}
\includegraphics[width=265pt]{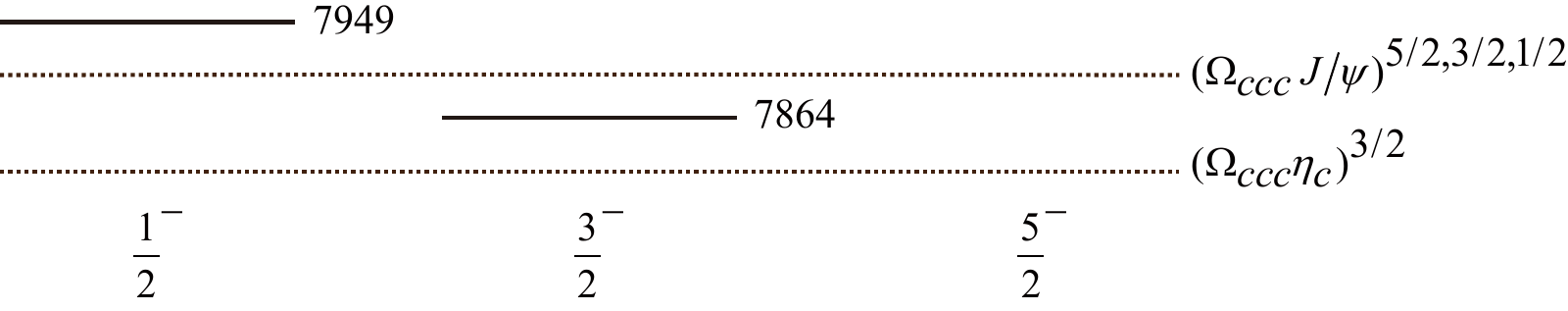}&
\includegraphics[width=265pt]{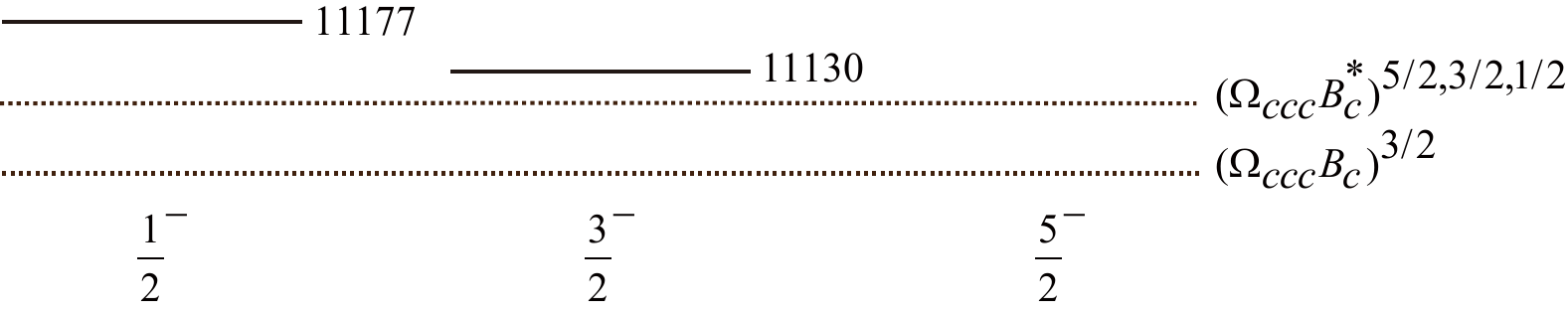}\\
(a) \begin{tabular}{c}  $cccc\bar{c}$ states\end{tabular} &(b)  $cccc\bar{b}$ states\\
\includegraphics[width=265pt]{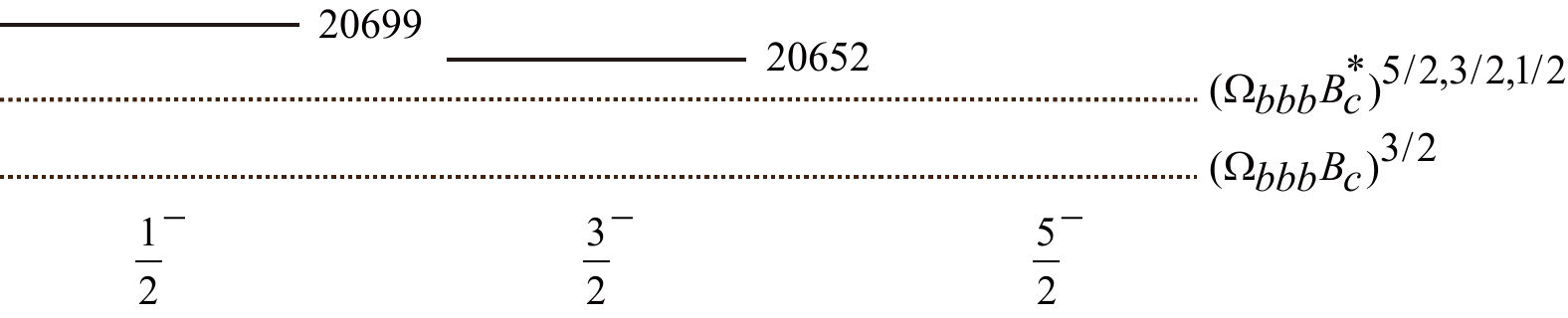}&
\includegraphics[width=265pt]{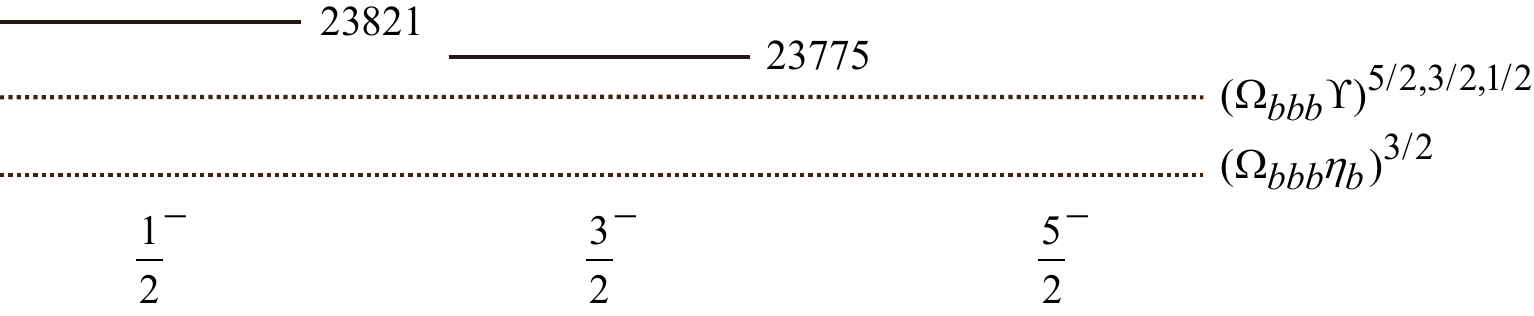}\\
(c) \begin{tabular}{c}  $bbbb\bar{c}$ states\end{tabular} &(d)  $bbbb\bar{b}$ states\\
\includegraphics[width=265pt]{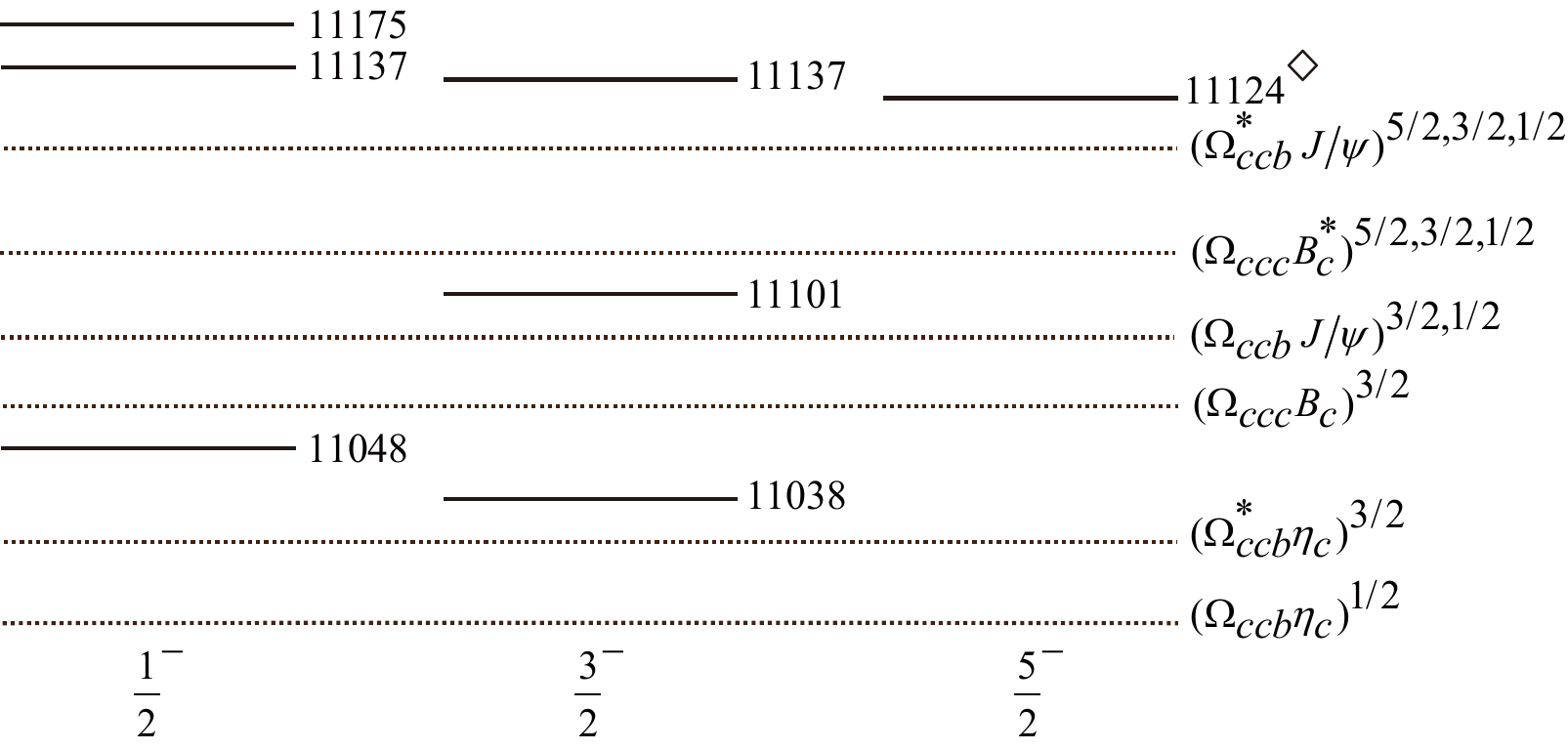}&
\includegraphics[width=265pt]{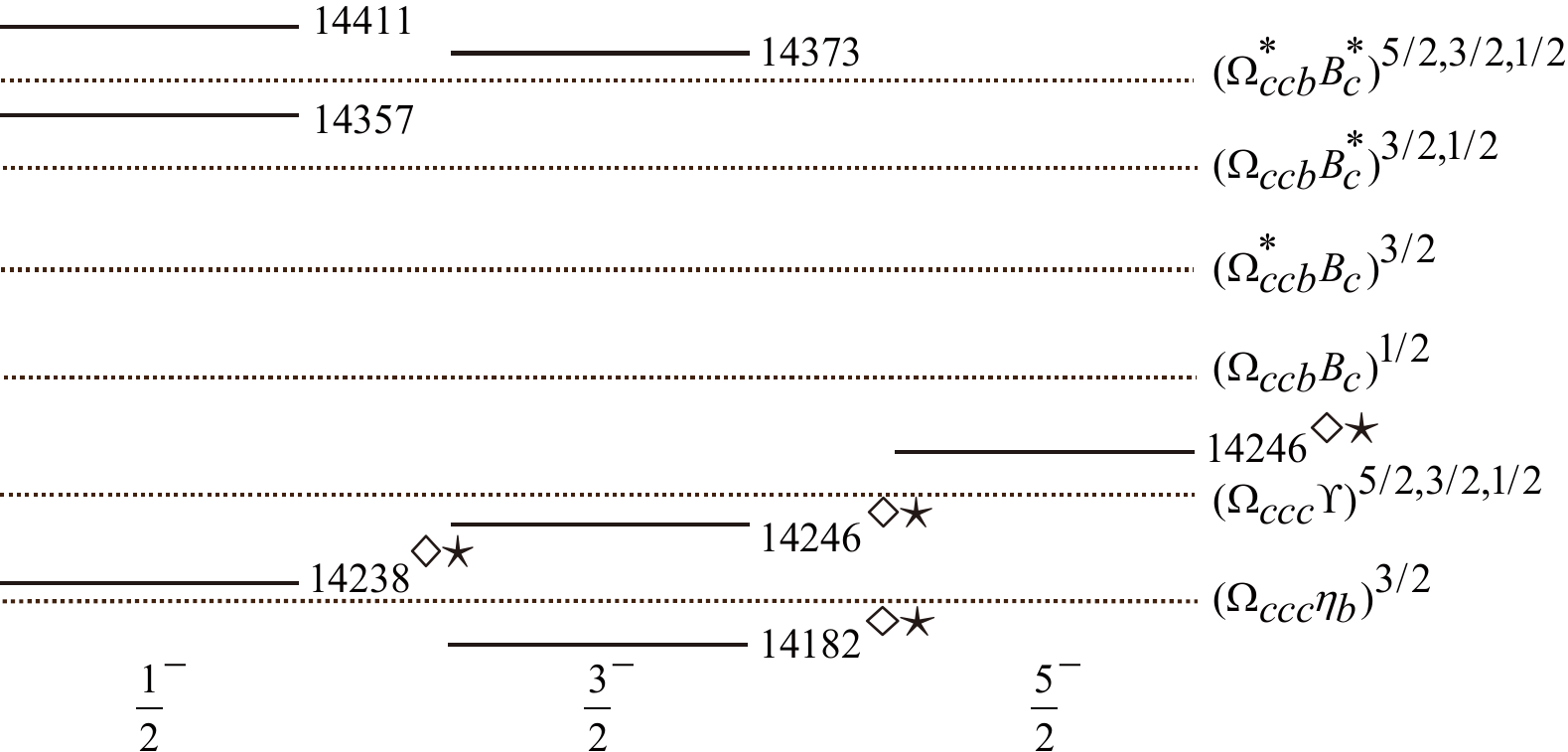}\\
(e) \begin{tabular}{c}  $cccb\bar{c}$ states\end{tabular} &(f)  $cccb\bar{b}$ states\\
\includegraphics[width=265pt]{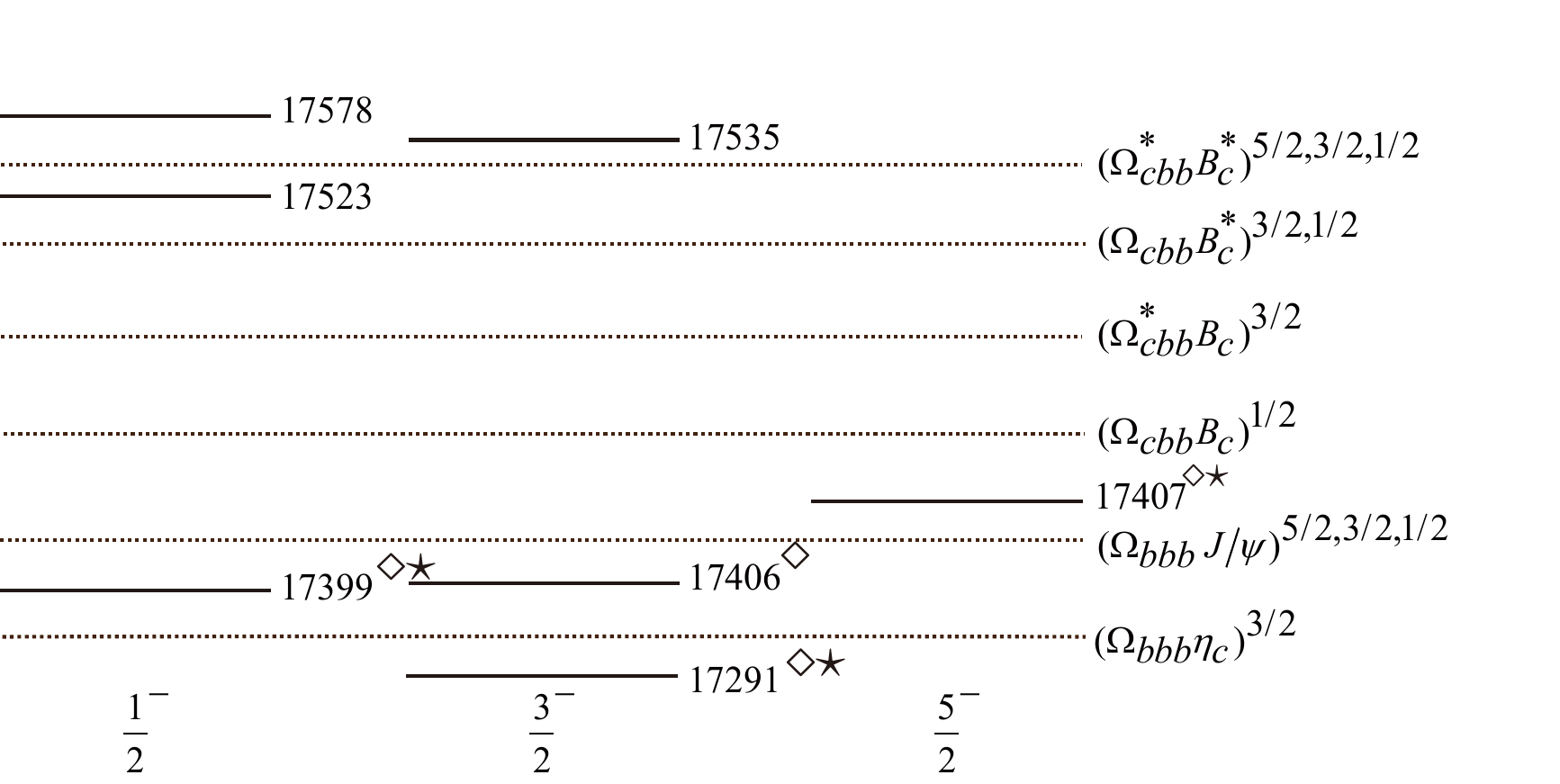}&
\includegraphics[width=265pt]{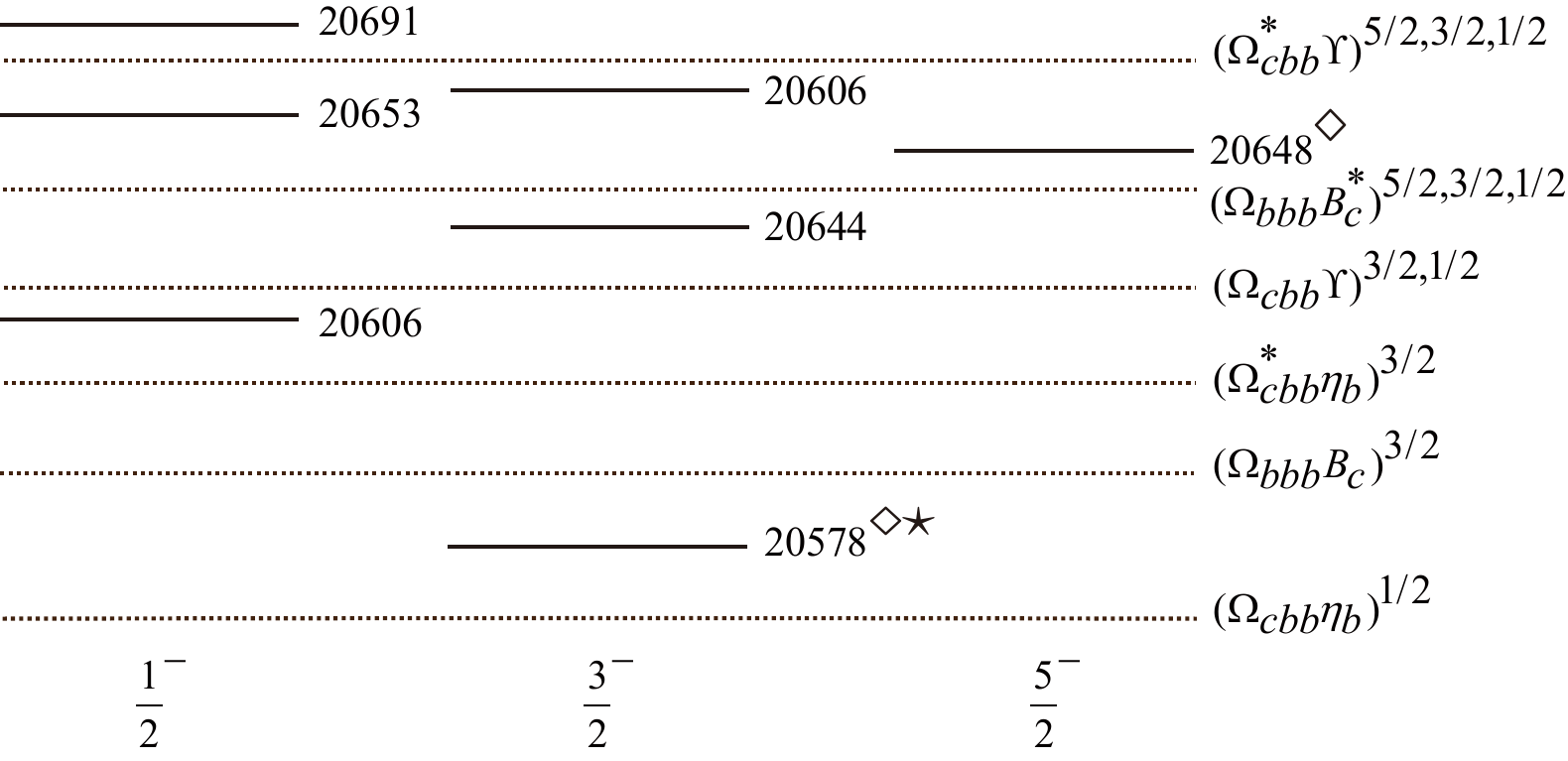}\\
(g) \begin{tabular}{c}  $bbbc\bar{c}$ states\end{tabular} &(h)  $bbbc\bar{b}$ states\\
\includegraphics[width=265pt]{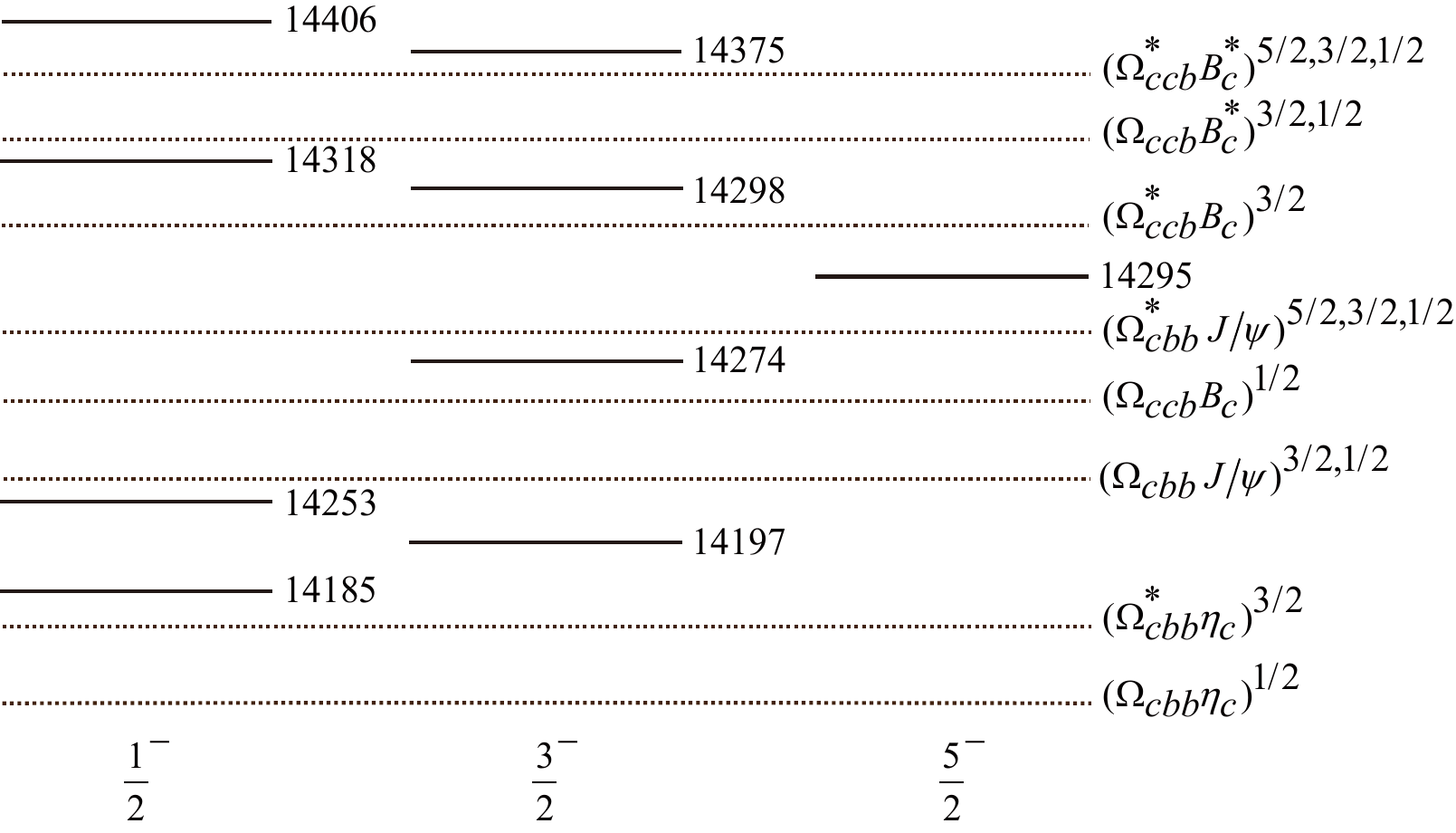}&
\includegraphics[width=265pt]{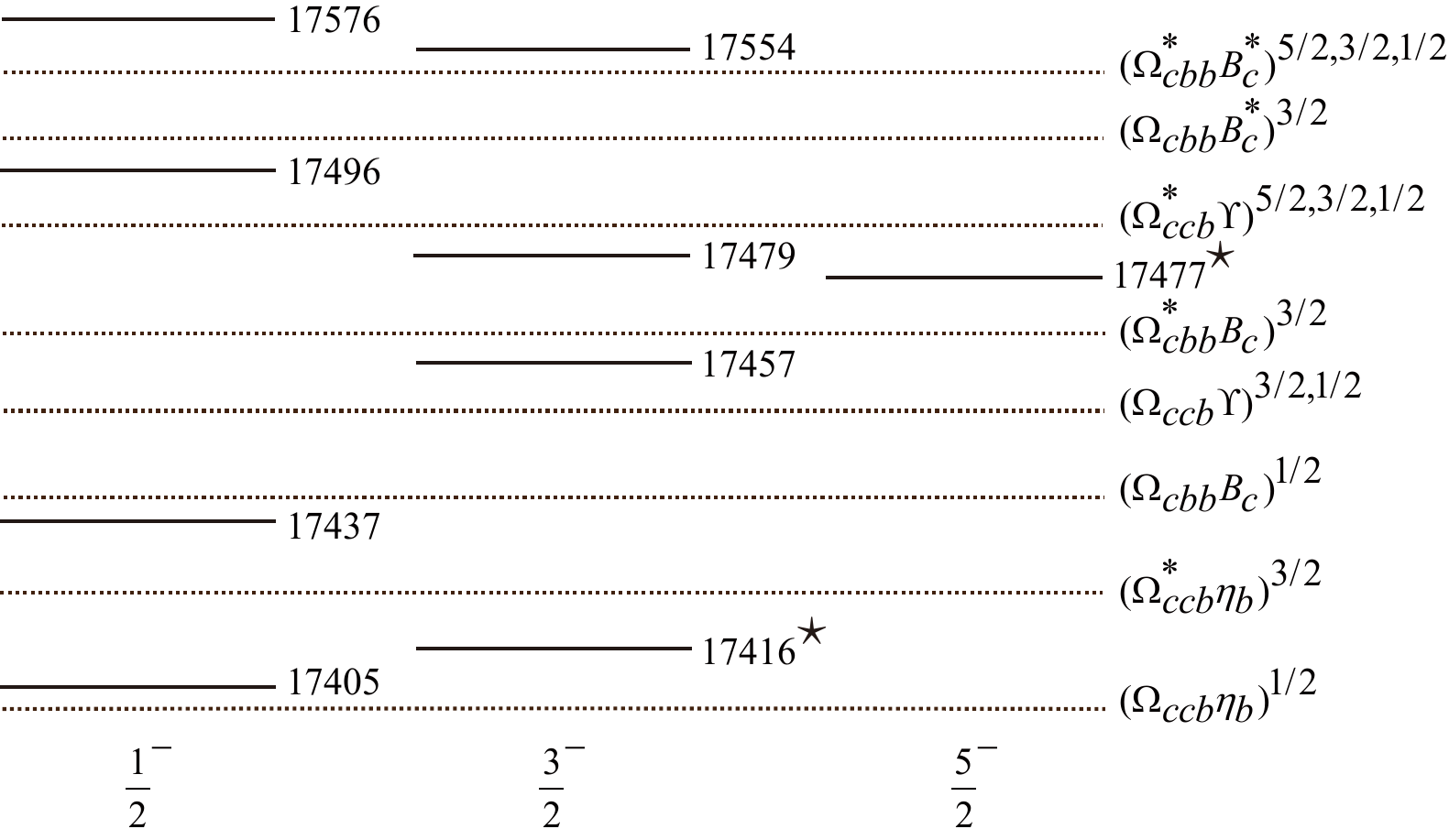}\\
(i) \begin{tabular}{c}  $ccbb\bar{c}$ states\end{tabular} &(j)  $ccbb\bar{b}$ states\\
\end{tabular}
\caption{
Relative positions (units: MeV) for the $cccc\bar{c}$, $cccc\bar{b}$, $bbbb\bar{c}$, $bbbb\bar{b}$, $cccb\bar{c}$, $cccb\bar{b}$, $bbbc\bar{c}$, $bbbc\bar{b}$, $ccbb\bar{c}$, and $ccbb\bar{b}$ pentaquark states labeled with solid lines.
The dotted lines denote various S-wave baryon-meson thresholds, and the superscripts of the labels, e.g. $(\Omega_{ccc}J/\psi)^{5/2,3/2,1/2}$, represent the possible total angular momenta of the channels.
We mark the relatively stable pentaquarks, unable to decay into the S-wave baryon-meson states, with ``$\star$" after their masses. We mark the pentaquark whose wave function overlaps with that of one special baryon-meson state more than 90\% with
``$\diamond$" after their masses. 
}\label{fig-QQQQQ}
\end{figure*}

The modified CMI model scheme takes the chromoelectric interaction explicitly compared to the reference mass scheme, and therefore we use the results in this scheme for the following analysis. According to the modified CMI model scheme, we present the masses of the $QQQQ\bar{Q}$ pentaquark states
and the relevant baryon-meson thresholds in Fig. \ref{fig-QQQQQ}. In Fig. \ref{fig-QQQQQ}, we label the possible total angular momenta of the S-wave baryon-meson states. When the spin of an initial pentaquark state is equal to the total angular momentum of the channel below, it may decay into that baryon-meson channel through the S wave.

Here we define the relatively "stable" pentaquarks as those which cannot decay into the S wave baryon-meson states. We label these stable pentaquark states with ``$\star$'' in the figure and tables.

In addition to the mass spectra, the eigenvectors of pentaquark states will also provide important information about the two-body strong decay of multiquark states \cite{Jaffe:1976ig,Strottman:1979qu,Weng:2019ynva,Weng:2020jao,Zhao:2014qva,Wang:2015epa}. Thus we calculate the overlaps of wave functions between a fully heavy pentaquark state and a particular baryon $\otimes$ meson state, and show them in Table \ref{eigenvector-QQQQQ}.

We can further study the decay of the fully heavy pentaquark states into the baryon $\otimes$ meson channels.
Here, we take the $cccc\bar{c}$ pentaquark states as an example to describe our calculation.
We transform the eigenvectors of the pentaquark states into the $ccc$ $\otimes$ $c\bar{c}$ configuration.
Normally, the $ccc$ and $c\bar{c}$ components inside the pentaquark can be either of color-singlet or of color-octet.
The former  can easily dissociate into an $S$-wave baryon and meson
[the so-called Okubo-Zweig-Iizuka-superallowed (OZI-superallowed] decays \cite{Jaffe:1976ig}),
while the latter cannot fall apart without the gluon exchange force.
For simplicity, in this work, we only focus on the OZI-superallowed pentaquark decay process.
The color singlet can be described by the direct product of a meson wave function and a baryon wave function.
For each decay mode, the branching fraction is proportional to the square of the coefficient of the corresponding component in the eigenvectors, and the strong decay phase space.

For the two-body decay via the $L$-wave process, the expression describing partial decay width can be parametrized as \cite{Weng:2019ynva,Weng:2020jao}
\begin{eqnarray}\label{Eq20}
\Gamma_{i}=\gamma_{i}\alpha\frac{k^{2L+1}}{m^{2L}}\cdot|c_{i}|^{2},
\end{eqnarray}
where $\alpha$ is an effective coupling constant,
$m$ is the mass of the initial state, and 
$k$ is the momentum of the final states in the rest frame. $c_{i}$ is the coefficient related to the corresponding baryon-meson component, which is the overlap of the wave functions shown in Table \ref{eigenvector-QQQQQ}. For the decay processes that we are interested in, $(k/m)^{2}$ is of $\mathcal{O}(10^{-2})$ or even smaller. Thus we only consider the $S$-wave decays.

$\gamma_{i}$ represents other factors that contribute to the decay widths $\Gamma_i$. For each process, $\Gamma_i$ also depends on the spatial wave functions of the initial
pentaquark state and the final meson and baryon.
In the quark model in the heavy quark limit, the spatial wave functions of the ground $S$-wave pseudoscalar and vector meson are the same \cite{Weng:2019ynva}. As a rough estimation, we introduce the following approximations to calculate the relative partial decay widths  of the pentaquark states:
\begin{eqnarray}\label{eq:gamma}
&&\gamma_{\Omega_{ccc}J/\psi}=\gamma_{\Omega_{ccc}\eta_{c}},\quad\gamma_{\Omega_{ccc}B_{c}^{*}}=\gamma_{\Omega_{ccc}B_{c}},\nonumber\\
&&\gamma_{\Omega_{bbb}B_{c}^{*}}=\gamma_{\Omega_{bbb}B_{c}},\quad\gamma_{\Omega_{bbb}\Upsilon}=\gamma_{\Omega_{bbb}\eta_{b}},\nonumber\\
&&\gamma_{\Omega_{ccc}B^{*}_{c}}=\gamma_{\Omega_{ccc}B_{c}},\quad\gamma_{\Omega_{ccc}\Upsilon}=\gamma_{\Omega_{ccc}\eta_{b}},\nonumber\\
&&\gamma_{\Omega_{bbb}J/\psi}=\gamma_{\Omega_{bbb}\eta_{c}}, \quad \gamma_{\Omega_{bbb}B^{*}_{c}}=\gamma_{\Omega_{bbb}B_{c}},
\nonumber\\
&&\gamma_{\Omega^{*}_{ccb}J/\psi}=\gamma_{\Omega^{*}_{ccb}\eta_{c}}=\gamma_{\Omega_{ccb}J/\psi}=\gamma_{\Omega_{ccb}\eta_{c}},
\nonumber\\
&&\gamma_{\Omega^{*}_{ccb}B^{*}_{c}}=\gamma_{\Omega^{*}_{ccb}B_{c}}=\gamma_{\Omega_{ccb}B^{*}_{c}}=\gamma_{\Omega_{ccb}B_{c}},
\nonumber\\
&&\gamma_{\Omega^{*}_{bbc}B^{*}_{c}}=\gamma_{\Omega^{*}_{bbc}B_{c}}=\gamma_{\Omega_{bbc}B^{*}_{c}}=\gamma_{\Omega_{bbc}B_{c}},
\nonumber\\
&&\gamma_{\Omega^{*}_{bbc}\Upsilon}=\gamma_{\Omega^{*}_{bbc}\eta_{b}}=\gamma_{\Omega_{bbc}\Upsilon}=\gamma_{\Omega_{bbc}\eta_{b}},
\nonumber\\&&
\gamma_{\Omega^{*}_{ccb}B^{*}_{c}}=\gamma_{\Omega^{*}_{ccb}B_{c}}=\gamma_{\Omega_{ccb}B^{*}_{c}}=\gamma_{\Omega_{ccb}B_{c}},
\nonumber\\&&
\gamma_{\Omega^{*}_{bbc}J/\psi}=\gamma_{\Omega^{*}_{bbc}}\eta_{c}=\gamma_{\Omega_{bbc}J/\psi}=\gamma_{\Omega_{bbc}}\eta_{c},
\nonumber\\&&
\gamma_{\Omega^{*}_{ccb}\Upsilon}=\gamma_{\Omega^{*}_{ccb}\eta_{b}}=\gamma_{\Omega_{ccb}\Upsilon}=\gamma_{\Omega_{ccb}\eta_{b}},
\nonumber\\&&
\gamma_{\Omega^{*}_{bbc}B^{*}_{c}}=\gamma_{\Omega^{*}_{bbc}B_{c}}=\gamma_{\Omega_{bbc}B^{*}_{c}}=\gamma_{\Omega_{bbc}B_{c}}.
\end{eqnarray}

We present $k\cdot|c_{i}|^{2}$ for each decay process in Table \ref{value-QQQQQ}. From Table \ref{value-QQQQQ}, one can roughly estimate the relative decay widths between different decay processes of different initial pentaquark states if neglecting the $\gamma_i$ differences.
{As a rough estimation about the ratios of the decay widths, we neglect the recoil momenta of quarks in the hadrons and thus use the approximation of Eq. (\ref{Eq20}). A complete five-body study will modify the phase space factor.}

Here, we divide the $QQQQ\bar{Q}$ pentaquark system into the following three groups:
\begin{enumerate}
\item[A.] The $cccc\bar{Q}$ and $bbbb\bar{Q}$ pentaquark subsystems;
\item[B.]  The $cccb\bar{Q}$ and $bbbc\bar{Q}$ pentaquark subsystems;
\item[C.]  The $ccbb\bar{Q}$ pentaquark subsystem.
\end{enumerate}
We discuss the mass spectra and strong decay properties of the $QQQQ\bar{Q}$ pentaquark system group by group.
For simplicity, we use $\rm P_{content}$(Mass, $I$, $J^{P}$) to label a specific pentaquark state.

\begin{table*}[b]
\centering \caption{The estimated masses for the $QQQQ\bar{Q}$ ($Q=c, b$) system in units of MeV. The eigenvalues of the $H_{\textrm{CMI}}$ matrix are listed in the second column.
The corresponding masses in the reference mass scheme, also labeled as the "1st scheme", are listed in third and/or fourth columns. The masses with the modified CMI model scheme, also labeled as the "2nd scheme" are presented in the last column.
}\label{mass-QQQQQ}
\renewcommand\arraystretch{1.25}
\begin{tabular}{cccc|c|cccc|c}
\bottomrule[1.5pt]
\bottomrule[0.5pt]
&\multicolumn{3}{c|}{1st scheme}&{2nd  scheme}&&\multicolumn{3}{c|}{1st scheme}&{2nd  scheme}\\
\bottomrule[1pt]
\multicolumn{5}{c|}{$cccc\bar{c}$}&\multicolumn{5}{c}{$cccc\bar{b}$}\\
$J^P$&Eigenvalue&($\Omega_{ccc}\eta_{c}$)&&Mass&$J^P$&Eigenvalue&$(\Omega_{ccc}B_{c})$&&Mass\\
\bottomrule[0.5pt]
$\frac{3}{2}^{-}$ &
$33.3$&
$7861$&
&
$7864$&
$\frac{3}{2}^{-}$ &
$44.0$&
$11131$&
&
$11130$\\
$\frac{1}{2}^{-}$ &
$118.1$&
$7946$&
&
$7949$&
$\frac{1}{2}^{-}$ &
$96.8$&
$11184$&
&
$11177$\\
\bottomrule[1pt]
\multicolumn{5}{c|}{$bbbb\bar{c}$}&\multicolumn{5}{c}{$bbbb\bar{b}$}\\
$J^P$&Eigenvalue&($\Omega_{bbb}B_{c}$)&&Mass&$J^P$&Eigenvalue&$(\Omega_{bbb}\eta_{b})$&&Mass\\
\bottomrule[0.5pt]
$\frac{3}{2}^{-}$ &
$16.0$&
$20639$&
&
$20652$&
$\frac{3}{2}^{-}$ &
$18.1$&
$23759$&
&
$23775$\\
$\frac{1}{2}^{-}$ &
$68.8$&
$20692$&
&
$20699$&
$\frac{1}{2}^{-}$ &
$64.5$&
$23805$&
&
$23821$\\
\bottomrule[1pt]
\multicolumn{5}{c|}{$cccb\bar{c}$}&\multicolumn{5}{c}{$cccb\bar{b}$}\\
$J^P$&Eigenvalue&($\Omega_{ccc}B_{c}$)&($\Omega_{ccb}\eta_{c}$)&Mass&$J^P$&Eigenvalue&$(\Omega_{ccc}\eta_{b})$&$(\Omega_{ccb}B_{c})$&Mass\\
\bottomrule[0.5pt]
$\frac{5}{2}^{-}$ &
$59.0$&
$11131$&
$10916$&
$11124$&
$\frac{5}{2}^{-}$ &
$41.9$&
$14247$&
$14172$&
$14246$\\
$\frac{3}{2}^{-}$ &
$\begin{pmatrix}59.0\\23.6\\-45.2\end{pmatrix}$&
$\begin{pmatrix}11146\\11111\\11042\end{pmatrix}$&
$\begin{pmatrix}10931\\10895\\10826\end{pmatrix}$&
$\begin{pmatrix}11137\\11101\\11038\end{pmatrix}$&
$\frac{3}{2}^{-}$ &
$\begin{pmatrix}47.1\\32.1\\-30.6\end{pmatrix}$&
$\begin{pmatrix}14252\\14237\\14174\end{pmatrix}$&
$\begin{pmatrix}14178\\14163\\14100\end{pmatrix}$&
$\begin{pmatrix}14373\\14246\\14182\end{pmatrix}$\\
$\frac{1}{2}^{-}$ &
$\begin{pmatrix}101.3\\61.2\\-26.2\end{pmatrix}$&
$\begin{pmatrix}11188\\11148\\11061\end{pmatrix}$&
$\begin{pmatrix}10973\\10933\\10845\end{pmatrix}$&
$\begin{pmatrix}11175\\11137\\11048\end{pmatrix}$&
$\frac{1}{2}^{-}$ &
$\begin{pmatrix}83.7\\45.5\\-8.8\end{pmatrix}$&
$\begin{pmatrix}14288\\14250\\14196\end{pmatrix}$&
$\begin{pmatrix}14214\\14176\\14122\end{pmatrix}$&
$\begin{pmatrix}14411\\14357\\14238\end{pmatrix}$\\
\bottomrule[1pt]
\multicolumn{5}{c|}{$bbbc\bar{c}$}&\multicolumn{5}{c}{$bbbc\bar{b}$}\\
$J^P$&Eigenvalue&($\Omega_{bbb}\eta_{c}$)&($\Omega_{bbc}B_{c}$)&Mass&$J^P$&Eigenvalue&$(\Omega_{bbb}B_{c})$&$(\Omega_{bbc}\eta_{b})$&Mass\\
\bottomrule[0.5pt]
$\frac{5}{2}^{-}$ &
$42.7$&
$17407$&
$17552$&
$17407$&
$\frac{5}{2}^{-}$ &
$32.0$&
$20655$&
$20659$&
$20648$\\
$\frac{3}{2}^{-}$ &
$\begin{pmatrix}44.7\\15.0\\-77.3\end{pmatrix}$&
$\begin{pmatrix}17409\\17379\\17287\end{pmatrix}$&
$\begin{pmatrix}17554\\17524\\17432\end{pmatrix}$&
$\begin{pmatrix}17535\\17406\\17291\end{pmatrix}$&
$\frac{3}{2}^{-}$ &
$\begin{pmatrix}34.0\\19.4\\-47.0\end{pmatrix}$&
$\begin{pmatrix}20657\\20642\\20576\end{pmatrix}$&
$\begin{pmatrix}20661\\20646\\20580\end{pmatrix}$&
$\begin{pmatrix}20654\\20644\\20578\end{pmatrix}$\\
$\frac{1}{2}^{-}$ &
$\begin{pmatrix}76.5\\36.2\\-18.3\end{pmatrix}$&
$\begin{pmatrix}17441\\17400\\17346\end{pmatrix}$&
$\begin{pmatrix}17586\\17545\\17491\end{pmatrix}$&
$\begin{pmatrix}17578\\17523\\17399\end{pmatrix}$&
$\frac{1}{2}^{-}$ &
$\begin{pmatrix}67.5\\29.6\\-18.7\end{pmatrix}$&
$\begin{pmatrix}20691\\20653\\20604\end{pmatrix}$&
$\begin{pmatrix}20694\\20657\\20608\end{pmatrix}$&
$\begin{pmatrix}20691\\20653\\20607\end{pmatrix}$\\
\bottomrule[1pt]
\multicolumn{5}{c|}{$ccbb\bar{c}$}&\multicolumn{5}{c}{$ccbb\bar{b}$}\\
$J^P$&Eigenvalue&($\Omega_{ccb}B_{c}$)&($\Omega_{bbc}\eta_{c}$)&Mass&$J^P$&Eigenvalue&$(\Omega_{ccb}\eta_{b})$&$(\Omega_{bbc}B_{c})$&Mass\\
\bottomrule[0.5pt]
$\frac{5}{2}^{-}$ &
$41.9$&
$14372$&
$14292$&
$14295$&
$\frac{5}{2}^{-}$ &
$35.5$&
$17484$&
$17545$&
$17477$\\
$\frac{3}{2}^{-}$ &
$\begin{pmatrix}52.2\\20.7\\16.9\\-63.0\end{pmatrix}$&
$\begin{pmatrix}14383\\14351\\14348\\14268\end{pmatrix}$&
$\begin{pmatrix}14302\\14271\\14267\\14187\end{pmatrix}$&
$\begin{pmatrix}14375\\14298\\14274\\14197\end{pmatrix}$&
$\frac{3}{2}^{-}$ &
$\begin{pmatrix}39.5\\24.0\\13.1\\-40.1\end{pmatrix}$&
$\begin{pmatrix}17488\\17472\\17461\\17408\end{pmatrix}$&
$\begin{pmatrix}17549\\17533\\17522\\17469\end{pmatrix}$&
$\begin{pmatrix}17554\\17479\\17457\\17416\end{pmatrix}$\\
$\frac{1}{2}^{-}$ &
$\begin{pmatrix}87.4\\46.5\\-19.1\\-69.6\end{pmatrix}$&
$\begin{pmatrix}14418\\14377\\14312\\14261\end{pmatrix}$&
$\begin{pmatrix}14338\\14297\\14231\\14181\end{pmatrix}$&
$\begin{pmatrix}14406\\14318\\14253\\14185\end{pmatrix}$&
$\frac{1}{2}^{-}$ &
$\begin{pmatrix}73.9\\35.8\\-15.5\\-49.0\end{pmatrix}$&
$\begin{pmatrix}17522\\17484\\17433\\17399\end{pmatrix}$&
$\begin{pmatrix}17583\\175454\\17494\\17460\end{pmatrix}$&
$\begin{pmatrix}17576\\17496\\17437\\17405\end{pmatrix}$\\
\bottomrule[0.5pt]
\bottomrule[1.5pt]
\end{tabular}
\end{table*}

\begin{table*}[t]
\centering \caption{The overlaps of wave functions between a fully heavy pentaquark state and a particular baryon $\otimes$ meson state. The masses are all in units of MeV.
See the caption of Fig. \ref{fig-QQQQQ} for meanings of ``$\diamond$" and ``$\star$".
}\label{eigenvector-QQQQQ}
\renewcommand\arraystretch{1.2}
\begin{tabular}{cl|cc|cccc|l|cc|cccc}
\bottomrule[1.5pt]
\bottomrule[0.5pt]
\multicolumn{2}{c|}{$cccc\bar{c}$}&\multicolumn{2}{c|}{$ccc\bigotimes c\bar{c}$}&\multicolumn{1}{l|}{$cccc\bar{b}$}&\multicolumn{2}{c|}{$ccc \bigotimes c\bar{b}$}&&\multicolumn{1}{l|}{$bbbb\bar{c}$}&\multicolumn{2}{c|}{$bbb\bigotimes b\bar{c}$}&\multicolumn{1}{l|}{$bbbb\bar{b}$}&\multicolumn{2}{c|}{$bbb \bigotimes b\bar{b}$}\\
$J^P$&Mass&$\Omega_{ccc}J/\psi$&$\Omega_{ccc}\eta_{c}$
&\multicolumn{1}{l|}{Mass}&$\Omega_{ccc} B_{c}^{*}$&\multicolumn{1}{c|}{$\Omega_{ccc} B_{c}$}&
&Mass&$\Omega_{bbb}B_{c}^{*}$&$\Omega_{bbb}B_{c}$
&\multicolumn{1}{l|}{Mass}&$\Omega_{bbb} \Upsilon$&\multicolumn{1}{c|}{$\Omega_{bbb} \eta_{b}$}\\
\Xcline{1-7}{0.7pt}\Xcline{9-14}{0.7pt}
$\frac{3}{2}^{-}$&7864&0.456&-0.354&\multicolumn{1}{c|}{11130}&0.456&\multicolumn{1}{c|}{-0.354}&&20652&0.456&-0.354&\multicolumn{1}{c|}{23775}&0.456&\multicolumn{1}{c|}{-0.354}\\
$\frac{1}{2}^{-}$&7949&-0.577&&\multicolumn{1}{c|}{11177}&0.577&\multicolumn{1}{c|}{}&&20699&-0.577&&\multicolumn{1}{c|}{23821}&0.577&\multicolumn{1}{c|}{}\\
\bottomrule[0.5pt]
\multicolumn{2}{c|}{$cccb\bar{c}$}&\multicolumn{2}{c}{$ccc\bigotimes b\bar{c}$}&\multicolumn{4}{c|}{$ccb\bigotimes c\bar{c}$}&\multicolumn{1}{l|}{$cccb\bar{b}$}&\multicolumn{2}{c}{$ccc\bigotimes b\bar{b}$}&\multicolumn{4}{c}{$ccb\bigotimes c\bar{b}$}\\
$J^P$&Mass&$\Omega_{ccc} B_{c}^{*}$&$\Omega_{ccc} B_{c}$
&$\Omega_{ccb}^{*}J/\psi$&$\Omega_{ccb}^{*}\eta_{c}$&$\Omega_{ccb}J/\psi$&$\Omega_{ccb}\eta_{c}$&
Mass&$\Omega_{ccc} \Upsilon$&$\Omega_{ccc} \eta_{b}$
&$\Omega_{ccb}^{*}B_{c}^{*}$&$\Omega_{ccb}^{*}B_{c}$&$\Omega_{ccb}B_{c}^{*}$&$\Omega_{ccb}B_{c}$\\
\bottomrule[0.7pt]
$\frac{5}{2}^{-}$&$11124\diamond$&1.000&&0.333&&&&$14246\diamond$&1.000&&0.333&&&\\
$\frac{3}{2}^{-}$&11137&0.812&0.236&-0.361&-0.008&0.275&&14373&-0.046&-0.120&0.521&-0.352&-0.209\\
&$11101$&0.569&-0.524&0.396&-0.279&0.102&&$14246\diamond$&0.999&-0.016&0.030&0.229&-0.242\\
&$11038$&0.130&0.818&0.095&-0.380&-0.250&&$14182\diamond\star$&0.011&0.993&0.154&0.214&0.215\\
$\frac{1}{2}^{-}$&11175&-0.543&&-0.587&&0.034&-0.001&14411&0.130&&-0.626&&0.199&0.087\\
&11137&-0.657&&0.172&&-0.519&-0.039&14357&-0.206&&0.126&&0.571&-0.297\\
&11048&0.523&&0.180&&0.316&-0.470&$14238\diamond\star$&0.969&&0.004&&0.068&0.355\\
\bottomrule[0.5pt]
\multicolumn{2}{c|}{$bbbc\bar{c}$}&\multicolumn{2}{c}{$bbb\bigotimes c\bar{c}$}&\multicolumn{4}{c|}{$bbc\bigotimes b\bar{c}$}&\multicolumn{1}{l|}{$bbbc\bar{b}$}&\multicolumn{2}{c}{$bbb\bigotimes c\bar{b}$}&\multicolumn{4}{c}{$bbc\bigotimes b\bar{b}$}\\
$J^P$&Mass&$\Omega_{bbb} J/\psi$&$\Omega_{bbb} \eta_{c}$
&$\Omega_{bbc}^{*}B_{c}^{*}$&$\Omega_{bbc}^{*}B_{c}$&$\Omega_{bbc}B_{c}^{*}$&$\Omega_{bbc}B_{c}$&
Mass&$\Omega_{bbb} B_{c}^{*}$&$\Omega_{bbb}B_{c}$
&$\Omega_{bbc}^{*}\Upsilon$&$\Omega_{bbc}^{*}\eta_{b}$&$\Omega_{bbc}\Upsilon$&$\Omega_{bbc}\eta_{b}$\\
\bottomrule[0.7pt]
$\frac{5}{2}^{-}$&$17407\diamond\star$&1.000&&0.333&&&&$20648\diamond$&1.000&&0.333&&&\\
$\frac{3}{2}^{-}$&17535&-0.045&-0.093&0.517&-0.358&-0.214&&20654&0.689&0.225&0.429&-0.079&-0.281\\
&$17406\diamond$&0.999&-0.009&-0.031&-0.231&0.240&&20644&0.724&-0.248&0.333&-0.362&0.084\\
&$17291\diamond\star$&0.005&0.996&-0.168&-0.203&-0.211&&$20578\diamond\star$&0.026&0.942&0.038&0.291&0.249\\
$\frac{1}{2}^{-}$&17578&-0.133&&-0.631&&0.169&0.106&20691&-0.505&&0.598&&-0.034&-0.019\\
&17523&-0.187&&0.092&&0.580&-0.296&20653&-0.662&&-0.156&&0.522&0.033\\
&$17399\diamond\star$&0.973&&0.010&&0.073&0.351&20607&0.554&&-0.158&&-0.311&0.470\\
\bottomrule[0.5pt]
\multicolumn{2}{c|}{$ccbb\bar{c}$}&\multicolumn{4}{c}{$bbc\bigotimes c\bar{c}$}&\multicolumn{4}{c|}{$ccb\bigotimes b\bar{c}$}\\
$J^P$&Mass&
$\Omega^{*}_{bbc}J/\psi$&\multicolumn{1}{c}{$\Omega^{*}_{bbc}\eta_{c}$}&$\Omega_{bbc}J/\psi$&\multicolumn{1}{c|}{$\Omega_{bbc}\eta_{c}$}&
$\Omega^{*}_{ccb}B_{c}^{*}$&\multicolumn{1}{c}{$\Omega^{*}_{ccb}B_{c}$}&\multicolumn{1}{c}{$\Omega_{ccb}B_{c}^{*}$}&\multicolumn{1}{c|}{$\Omega_{ccb}B_{c}$}\\
\Xcline{1-10}{0.7pt}
$\frac52^{-}$&14295&-0.577&\multicolumn{1}{c}{}&&\multicolumn{1}{c|}{}&-0.577&\multicolumn{1}{c}{}&\multicolumn{1}{c}{}&\multicolumn{1}{c|}{}\\
$\frac32^{-}$
&14375&-0.070&\multicolumn{1}{c}{-0.088}&-0.019&\multicolumn{1}{c|}{}&0.592&\multicolumn{1}{c}{-0.316}&\multicolumn{1}{c}{0.450}&\multicolumn{1}{c|}{}\\
&14298&0.631&\multicolumn{1}{c}{-0.057}&0.370&\multicolumn{1}{c|}{}&-0.138&\multicolumn{1}{c}{-0.325}&\multicolumn{1}{c}{0.056}&\multicolumn{1}{c|}{}\\
&14274&-0.251&\multicolumn{1}{c}{0.126}&0.507&\multicolumn{1}{c|}{}&-0.224&\multicolumn{1}{c}{0.242}&\multicolumn{1}{c}{0.471}&\multicolumn{1}{c|}{}\\
&14197&-0.074&\multicolumn{1}{c}{0.624}&-0.225&\multicolumn{1}{c|}{}&0.229&\multicolumn{1}{c}{0.390}&\multicolumn{1}{c}{-0.128}&\multicolumn{1}{c|}{}\\
$\frac12^{-}$
&14406&0.202&\multicolumn{1}{c}{}&-0.074&\multicolumn{1}{c|}{-0.025}&-0.722&\multicolumn{1}{c}{}&\multicolumn{1}{c}{-0.200}&\multicolumn{1}{c|}{-0.242}\\
&14318&0.666&\multicolumn{1}{c}{}&0.277&\multicolumn{1}{c|}{0.033}&-0.008&\multicolumn{1}{c}{}&\multicolumn{1}{c}{-0.348}&\multicolumn{1}{c|}{-0.156}\\
&14253&0.225&\multicolumn{1}{c}{}&-0.505&\multicolumn{1}{c|}{0.120}&-0.012&\multicolumn{1}{c}{}&\multicolumn{1}{c}{0.273}&\multicolumn{1}{c|}{-0.522}\\
&14185&0.146&\multicolumn{1}{c}{}&0.155&\multicolumn{1}{c|}{0.633}&0.184&\multicolumn{1}{c}{}&\multicolumn{1}{c}{-0.354}&\multicolumn{1}{c|}{-0.248}\\
\Xcline{1-10}{0.5pt}
\multicolumn{2}{c|}{$ccbb\bar{b}$}&\multicolumn{4}{c}{$bbc\bigotimes c\bar{b}$}&\multicolumn{4}{c|}{$ccb\bigotimes b\bar{b}$}\\
$J^P$&Mass&
$\Omega^{*}_{bbc}B_{c}^{*}$&\multicolumn{1}{c}{$\Omega^{*}_{bbc}B_{c}$}&$\Omega_{bbc}B_{c}^{*}$&\multicolumn{1}{c|}{$\Omega_{bbc}B_{c}$}&
$\Omega^{*}_{ccb}\Upsilon$&\multicolumn{1}{c}{$\Omega^{*}_{ccb}\eta_{b}$}&\multicolumn{1}{c}{$\Omega_{ccb}\Upsilon$}&\multicolumn{1}{c|}{$\Omega_{ccb}\eta_{b}$}\\
\Xcline{1-10}{0.7pt}
$\frac52^{-}$
&$17477\star$&-0.577&\multicolumn{1}{c}{}&&\multicolumn{1}{c|}{}&-0.577&\multicolumn{1}{c}{}&\multicolumn{1}{c}{}&\multicolumn{1}{c|}{}\\
$\frac32^{-}$
&17554&0.601&\multicolumn{1}{c}{-0.308}&0.440&\multicolumn{1}{c|}{}&-0.052&\multicolumn{1}{c}{-0.117}&\multicolumn{1}{c}{-0.008}&\multicolumn{1}{c|}{}\\
&17479&-0.082&\multicolumn{1}{c}{-0.289}&-0.051&\multicolumn{1}{c|}{}&0.669&\multicolumn{1}{c}{-0.166}&\multicolumn{1}{c}{0.315}&\multicolumn{1}{c|}{}\\
&17457&0.265&\multicolumn{1}{c}{-0.123}&-0.492&\multicolumn{1}{c|}{}&-0.144&\multicolumn{1}{c}{0.193}&\multicolumn{1}{c}{0.530}&\multicolumn{1}{c|}{}\\
&$17416\star$&-0.184&\multicolumn{1}{c}{-0.472}&0.076&\multicolumn{1}{c|}{}&0.043&\multicolumn{1}{c}{0.581}&\multicolumn{1}{c}{-0.253}&\multicolumn{1}{c|}{}\\
$\frac12^{-}$
&17576&-0.723&\multicolumn{1}{c}{}&-0.195&\multicolumn{1}{c|}{-0.226}&0.219&\multicolumn{1}{c}{}&\multicolumn{1}{c}{-0.080}&\multicolumn{1}{c|}{-0.026}\\
&17496&0.038&\multicolumn{1}{c}{}&0.317&\multicolumn{1}{c|}{0.150}&0.675&\multicolumn{1}{c}{}&\multicolumn{1}{c}{0.281}&\multicolumn{1}{c|}{0.088}\\
&17437&0.069&\multicolumn{1}{c}{}&-0.365&\multicolumn{1}{c|}{0.438}&-0.224&\multicolumn{1}{c}{}&\multicolumn{1}{c}{0.458}&\multicolumn{1}{c|}{-0.277}\\
&17405&-0.164&\multicolumn{1}{c}{}&0.300&\multicolumn{1}{c|}{0.389}&0.036&\multicolumn{1}{c}{}&\multicolumn{1}{c}{0.257}&\multicolumn{1}{c|}{0.576}\\
\bottomrule[0.5pt]
\bottomrule[1.5pt]
\end{tabular}
\end{table*}

\begin{table*}[t]
\centering \caption{The values of $k\cdot |c_{i}|^{2}$ for the $cccc\bar{c}$, $cccc\bar{b}$, $bbbb\bar{c}$, $bbbb\bar{b}$, $cccb\bar{c}$, $cccb\bar{b}$, $bbbc\bar{c}$, $bbbc\bar{b}$, $ccbb\bar{c}$, and $ccbb\bar{b}$ pentaquark states. The masses are all in units of MeV. The decay channel is marked with ``$\times$'' if kinetically forbidden.
See the caption of Fig. \ref{fig-QQQQQ} for meanings of ``$\diamond$" and ``$\star$".
One can roughly estimate the relative decay widths between different decay processes of different initial pentaquark states with this table if neglecting the $\gamma_i$ differences.
}\label{value-QQQQQ}
\renewcommand\arraystretch{1.2}
\begin{tabular}{cl|cc|cccc|l|cc|cccc}
\bottomrule[1.5pt]
\bottomrule[0.5pt]
&$cccc\bar{c}$&\multicolumn{2}{c|}{$ccc\bigotimes c\bar{c}$}&\multicolumn{1}{l|}{$cccc\bar{b}$}&\multicolumn{2}{c|}{$ccc \bigotimes c\bar{b}$}&&$bbbb\bar{c}$&\multicolumn{2}{c|}{$bbb\bigotimes b\bar{c}$}&\multicolumn{1}{l|}{$bbbb\bar{b}$}&\multicolumn{2}{c|}{$bbb \bigotimes b\bar{b}$}\\
$J^P$&Mass&$\Omega_{ccc}J/\psi$&$\Omega_{ccc}\eta_{c}$
&\multicolumn{1}{l|}{Mass}&$\Omega_{ccc} B_{c}^{*}$&\multicolumn{1}{c|}{$\Omega_{ccc} B_{c}$}&
&\multicolumn{1}{l|}{Mass}&$\Omega_{bbb}B_{c}^{*}$&$\Omega_{bbb}B_{c}$
&\multicolumn{1}{l|}{Mass}&$\Omega_{bbb} \Upsilon$&\multicolumn{1}{c|}{$\Omega_{bbb} \eta_{b}$}\\
\Xcline{1-7}{0.7pt}\Xcline{9-14}{0.7pt}
$\frac{3}{2}^{-}$&7864&$\times$&74&\multicolumn{1}{c|}{11130}&$\times$&\multicolumn{1}{c|}{77}&&20652&39&96&\multicolumn{1}{c|}{23775}&49&\multicolumn{1}{c|}{108}\\
$\frac{1}{2}^{-}$&7949&167&&\multicolumn{1}{c|}{11177}&181&\multicolumn{1}{c|}{}&&20699&224&&\multicolumn{1}{c|}{23821}&254&\multicolumn{1}{c|}{}\\
\bottomrule[0.5pt]
\multicolumn{2}{c|}{$cccb\bar{c}$}&\multicolumn{2}{c}{$ccc\bigotimes b\bar{c}$}&\multicolumn{4}{c|}{$ccb\bigotimes c\bar{c}$}&$cccb\bar{b}$&\multicolumn{2}{c}{$ccc\bigotimes b\bar{b}$}&\multicolumn{4}{c}{$ccb\bigotimes c\bar{b}$}\\
$J^P$&Mass&$\Omega_{ccc} B_{c}^{*}$&$\Omega_{ccc} B_{c}$
&$\Omega_{ccb}^{*}J/\psi$&$\Omega_{ccb}^{*}\eta_{c}$&$\Omega_{ccb}J/\psi$&$\Omega_{ccb}\eta_{c}$&
Mass&$\Omega_{ccc} \Upsilon$&$\Omega_{ccc} \eta_{b}$
&$\Omega_{ccb}^{*}B_{c}^{*}$&$\Omega_{ccb}^{*}B_{c}$&$\Omega_{ccb}B_{c}^{*}$&$\Omega_{ccb}B_{c}$\\
\bottomrule[0.7pt]
$\frac{5}{2}^{-}$
&$11124\diamond$&14&&$\times$&&&&$14246\diamond$&15&&$\times$&&&\\
$\frac{3}{2}^{-}$
&11137&177&36&37&0.05&36&&14373&2&16&82&91&25\\
&$11101$&$\times$&130&$\times$&50&3&&$14246\diamond$&$\times$&0.2&$\times$&$\times$&$\times$\\
&$11038$&$\times$&$\times$&$\times$&54&$\times$&&$14182\diamond\star$&$\times$&$\times$&$\times$&$\times$&$\times$\\
$\frac{1}{2}^{-}$
&11175&156&&173&&0.7&0.002&14411&18&&236&&30&8\\
&11137&117&&9&&127&1&14357&36&&$\times$&&147&71\\
&11048&$\times$&&$\times$&&$\times$&125&$14238\diamond\star$&$\times$&&$\times$&&$\times$&$\times$\\
\bottomrule[0.5pt]
\multicolumn{2}{c|}{$bbbc\bar{c}$}&\multicolumn{2}{c}{$bbb\bigotimes c\bar{c}$}&\multicolumn{4}{c|}{$bbc\bigotimes b\bar{c}$}&\multicolumn{1}{l|}{$bbbc\bar{b}$}&\multicolumn{2}{c}{$bbb\bigotimes c\bar{b}$}&\multicolumn{4}{c}{$bbc\bigotimes b\bar{b}$}\\
$J^P$&Mass&$\Omega_{bbb} J/\psi$&$\Omega_{bbb} \eta_{c}$
&$\Omega_{bbc}^{*}B_{c}^{*}$&$\Omega_{bbc}^{*}B_{c}$&$\Omega_{bbc}B_{c}^{*}$&$\Omega_{bbc}B_{c}$&
Mass&$\Omega_{bbb} B_{c}^{*}$&$\Omega_{bbb}B_{c}$
&$\Omega_{bbc}^{*}\Upsilon$&$\Omega_{bbc}^{*}\eta_{b}$&$\Omega_{bbc}\Upsilon$&$\Omega_{bbc}\eta_{b}$\\
\bottomrule[0.7pt]
$\frac{5}{2}^{-}$
&$17407\diamond\star$&$\times$&&$\times$&&&&$20648\diamond$&77&&$\times$&&&\\
$\frac{3}{2}^{-}$
&17535&2&10&24&92&24&&20654&110&40&$\times$&5&43\\
&$17406\diamond$&$\times$&0.06&$\times$&$\times$&$\times$&&20644&$\times$&44&$\times$&92&3\\
&$17291\diamond\star$&$\times$&$\times$&$\times$&$\times$&$\times$&&$20578\diamond\star$&$\times$&$\times$&$\times$&$\times$&$\times$\\
$\frac{1}{2}^{-}$
&17578&17&&238&&22&12&20691&158&&213&&0.9&0.4\\
&17523&27&&$\times$&&134&72&20653&93&&$\times$&&145&1\\
&$17399\diamond\star$&$\times$&&$\times$&&$\times$&$\times$&20607&$\times$&&$\times$&&$\times$&146\\
\bottomrule[0.5pt]
\multicolumn{2}{c|}{$ccbb\bar{c}$}&\multicolumn{4}{c}{$bbc\bigotimes c\bar{c}$}&\multicolumn{4}{c|}{$ccb\bigotimes b\bar{c}$}\\
$J^P$&Mass&
$\Omega^{*}_{bbc}J/\psi$&\multicolumn{1}{c}{$\Omega^{*}_{bbc}\eta_{c}$}&$\Omega_{bbc}J/\psi$&\multicolumn{1}{c|}{$\Omega_{bbc}\eta_{c}$}&
$\Omega^{*}_{ccb}B_{c}^{*}$&\multicolumn{1}{c}{$\Omega^{*}_{ccb}B_{c}$}&\multicolumn{1}{c}{$\Omega_{ccb}B_{c}^{*}$}&\multicolumn{1}{c|}{$\Omega_{ccb}B_{c}$}\\
\Xcline{1-10}{0.7pt}
$\frac52^{-}$&14295&29&\multicolumn{1}{c}{}&&\multicolumn{1}{c|}{}&0&\multicolumn{1}{c}{}&\multicolumn{1}{c}{}&\multicolumn{1}{c|}{}\\
$\frac32^{-}$
&14375&3&\multicolumn{1}{c}{8}&0.3&\multicolumn{1}{c|}{}&115&\multicolumn{1}{c}{74}&\multicolumn{1}{c}{117}&\multicolumn{1}{c|}{}\\
&14298&61&\multicolumn{1}{c}{2}&57&\multicolumn{1}{c|}{}&$\times$&\multicolumn{1}{c}{10}&\multicolumn{1}{c}{$\times$}&\multicolumn{1}{c|}{}\\
&14274&$\times$&\multicolumn{1}{c}{11}&63&\multicolumn{1}{c|}{}&$\times$&\multicolumn{1}{c}{$\times$}&\multicolumn{1}{c}{$\times$}&\multicolumn{1}{c|}{}\\
&14197&$\times$&\multicolumn{1}{c}{109}&$\times$&\multicolumn{1}{c|}{}&$\times$&\multicolumn{1}{c}{$\times$}&\multicolumn{1}{c}{$\times$}&\multicolumn{1}{c|}{}\\
$\frac12^{-}$
&14406&30&\multicolumn{1}{c}{}&5&\multicolumn{1}{c|}{0.7}&298&\multicolumn{1}{c}{}&\multicolumn{1}{c}{30}&\multicolumn{1}{c|}{58}\\
&14318&153&\multicolumn{1}{c}{}&40&\multicolumn{1}{c|}{1}&$\times$&\multicolumn{1}{c}{}&\multicolumn{1}{c}{$\times$}&\multicolumn{1}{c|}{15}\\
&14253&$\times$&\multicolumn{1}{c}{}&$\times$&\multicolumn{1}{c|}{10}&$\times$&\multicolumn{1}{c}{}&\multicolumn{1}{c}{$\times$}&\multicolumn{1}{c|}{$\times$}\\
&14185&$\times$&\multicolumn{1}{c}{}&$\times$&\multicolumn{1}{c|}{166}&$\times$&\multicolumn{1}{c}{}&\multicolumn{1}{c}{$\times$}&\multicolumn{1}{c|}{$\times$}\\
\Xcline{1-10}{0.5pt}
\multicolumn{2}{c|}{$ccbb\bar{b}$}&\multicolumn{4}{c}{$bbc\bigotimes c\bar{b}$}&\multicolumn{4}{c|}{$ccb\bigotimes b\bar{b}$}\\
$J^P$&Mass&
$\Omega^{*}_{bbc}B_{c}^{*}$&\multicolumn{1}{c}{$\Omega^{*}_{bbc}B_{c}$}&$\Omega_{bbc}B_{c}^{*}$&\multicolumn{1}{c|}{$\Omega_{bbc}B_{c}$}&
$\Omega^{*}_{ccb}\Upsilon$&\multicolumn{1}{c}{$\Omega^{*}_{ccb}\eta_{b}$}&\multicolumn{1}{c}{$\Omega_{ccb}\Upsilon$}&\multicolumn{1}{c|}{$\Omega_{ccb}\eta_{b}$}\\
\Xcline{1-10}{0.7pt}
$\frac52^{-}$
&$17477\star$&$\times$&\multicolumn{1}{c}{}&&\multicolumn{1}{c|}{}&$\times$&\multicolumn{1}{c}{}&\multicolumn{1}{c}{}&\multicolumn{1}{c|}{}\\
$\frac32^{-}$
&17554&145&\multicolumn{1}{c}{78}&125&\multicolumn{1}{c|}{}&2&\multicolumn{1}{c}{15}&\multicolumn{1}{c}{0.06}&\multicolumn{1}{c|}{}\\
&17479&$\times$&\multicolumn{1}{c}{21}&$\times$&\multicolumn{1}{c|}{}&$\times$&\multicolumn{1}{c}{20}&\multicolumn{1}{c}{49}&\multicolumn{1}{c|}{}\\
&17457&$\times$&\multicolumn{1}{c}{$\times$}&$\times$&\multicolumn{1}{c|}{}&$\times$&\multicolumn{1}{c}{21}&\multicolumn{1}{c}{68}&\multicolumn{1}{c|}{}\\
&$17416\star$&$\times$&\multicolumn{1}{c}{$\times$}&$\times$&\multicolumn{1}{c|}{}&$\times$&\multicolumn{1}{c}{$\times$}&\multicolumn{1}{c}{$\times$}&\multicolumn{1}{c|}{}\\
$\frac12^{-}$
&17576&303&\multicolumn{1}{c}{}&29&\multicolumn{1}{c|}{54}&43&\multicolumn{1}{c}{}&\multicolumn{1}{c}{7}&\multicolumn{1}{c|}{0.9}\\
&17496&$\times$&\multicolumn{1}{c}{}&$\times$&\multicolumn{1}{c|}{15}&160&\multicolumn{1}{c}{}&\multicolumn{1}{c}{50}&\multicolumn{1}{c|}{7}\\
&17437&$\times$&\multicolumn{1}{c}{}&$\times$&\multicolumn{1}{c|}{$\times$}&$\times$&\multicolumn{1}{c}{}&\multicolumn{1}{c}{$\times$}&\multicolumn{1}{c|}{49}\\
&17405&$\times$&\multicolumn{1}{c}{}&$\times$&\multicolumn{1}{c|}{$\times$}&$\times$&\multicolumn{1}{c}{}&\multicolumn{1}{c}{$\times$}&\multicolumn{1}{c|}{122}\\
\bottomrule[0.5pt]
\bottomrule[1.5pt]
\end{tabular}
\end{table*}


\subsection{The $cccc\bar{Q}$ and $bbbb\bar{Q}$ pentaquark states}
We first discuss the fully heavy pentaquark states with $cccc\bar{c}$, $cccc\bar{b}$, $bbbb\bar{c}$, and $bbbb\bar{b}$ flavor configurations.
The $cccc\bar{b}$ and $bbbb\bar{c}$ states are the absolute exotic states which have the different flavor quantum numbers from the conventional baryons.
Because of the strong symmetrical constraint from the Pauli principle, i.e., fully antisymmetric among the first four charm quarks,
we only find two $cccc\bar{c}$ states: an $I(J^{P})=0(3/2^{-})$ state, $\rm P_{c^{4}\bar{c}}$(7864, $0$, $3/2^{-}$), and an $I(J^{P})=0(1/2^{-})$ state, $\rm P_{c^{4}\bar{c}}$(7949, $0$, $1/2^{-}$).
Similarly, there are also only two pentaquark states in the $cccc\bar{b}$, $bbbb\bar{c}$, and $bbbb\bar{b}$ subsystems.

From Fig. \ref{fig-QQQQQ} (a)-(d), the $J^{P}=3/2^{-}$ states generally have smaller masses than the $J^{P}=1/2^{-}$ states in the $cccc\bar{Q}$ and $bbbb\bar{Q}$ pentaquark subsystems.
Meanwhile, also from Fig. \ref{fig-QQQQQ} (a)-(d), the masses of all the $cccc\bar{Q}$ and $bbbb\bar{Q}$ pentaquark states are larger than the thresholds of the lowest possible baryon-meson systems. Thus, no stable pentaquark state exists in the $cccc\bar{Q}$ and $bbbb\bar{Q}$ pentaquark subsystems. The lowest baryon-meson channels are their dominant decay modes.
In the future, searching for exotic signals in these baryon-meson strong decay channels would be an interesting topic.

The $cccc\bar{c}$ subsystem has one decay mode:
$ccc\otimes c\bar{c}$, which could be $\Omega_{ccc}J/\psi$ or $\Omega_{ccc}\eta_{c}$. However, each $cccc\bar{c}$ state has only one decay channel from Table \ref{value-QQQQQ}. The $J^{P}=1/2^{-}$ $cccc\bar{c}$ pentaquark state cannot decay into S-wave $\Omega_{ccc}\eta_{c}$ because of the constraint of angular conservation law, while the $J^{P}=3/2^{-}$ one cannot decay into $\Omega_{ccc}J/\psi$ since the mass is below the threshold.

From Table \ref{value-QQQQQ}, the ratio of decay widths between branching channels for $\rm P_{b^4\bar{c}}(20652,0,3/2^-)$ with the assumptions of Eq. (\ref{eq:gamma}) is
\begin{equation}
\Gamma_{\Omega_{bbb}B_{c}^{*}}:\Gamma_{\Omega_{bbb}B_{c}}=0.4:1,
\end{equation}
and for $\rm P_{b^4\bar{b}}(23775,0,3/2^{-})$ it is
\begin{equation}
\Gamma_{\Omega_{bbb}\Upsilon}:\Gamma_{\Omega_{bbb}\eta_{b}}=0.4:1.
\end{equation}
From the above ratios, one notices that $\rm P_{b^4\bar{c}}(20652,0,3/2^-)$ and $\rm P_{b^4\bar{b}}(23775,0,3/2^{-})$ have very similar decay behaviors: the $\Omega_{bbb}B_{c}^{*}$, $\Omega_{bbb}B_{c}$, $\Omega_{bbb}\Upsilon$, and $\Omega_{bbb}\eta_{b}$ are their dominant decay channels.

\subsection{The $cccb\bar{Q}$ and $bbbc\bar{Q}$ pentaquark states}\label{secw}
For the $cccb\bar{c}$, $cccb\bar{b}$, $bbbc\bar{c}$, and $bbbc\bar{b}$ pentaquark states, each has six possible strong decay channels from Fig. \ref{fig-QQQQQ} (e)-(h). For example, the $cccb\bar{c}$ subsystem may decay into $\Omega^{*}_{ccb}J/\psi$, $\Omega_{ccc}B_{c}^{*}$,
$\Omega_{ccb}J/\psi$, $\Omega_{ccc}B_{c}$, $\Omega^{*}_{ccb}\eta_{c}$, and $\Omega_{ccb}\eta_{c}$.

According to Table \ref{eigenvector-QQQQQ}, the $I(J^{P})=0(5/2^{-})$ $cccb\bar{b}$ pentaquark state $\rm P_{c^{3}b\bar{b}}(14246, 0, 5/2^{-})$ couples completely to the $\Omega_{ccc}\Upsilon$ system, which can be written as a direct product of a baryon $\Omega_{ccc}$ and a meson $\Upsilon$. Meanwhile, the $\rm P_{c^{3}b\bar{b}}(14246, 0, 3/2^{-})$, $\rm P_{c^{3}b\bar{b}}(14182, 0, 3/2^{-})$ , and $\rm P_{c^{3}b\bar{b}}(14238, 0, 1/2^{-})$ states couple almost completely to the $\Omega_{ccc}\Upsilon$, $\Omega_{ccc}\eta_{b}$, and $\Omega_{ccc}\Upsilon$ baryon-meson systems, respectively. This kind of pentaquark behaves similarly to the ordinary scattering state made of a baryon and meson if the inner interaction is not strong, but could also be a resonance or bound state dynamically generated by the baryon and a meson with a strong interaction. These kinds of pentaquarks deserve a more careful study with some hadron-hadron interaction models in the future. We label these states with $``\diamond"$ in the figure and tables. 

{
Moreover, $1 \otimes 1$ components in the $cccb\bar b$ subsystem are dominant and thus the lowest two pentaquarks are closed to the corresponding thresholds $\Omega_{ccc}$+$\Upsilon$ and $\Omega_{ccc}$+$\eta_{b}$, respectively. It is the attractive interaction between $\Omega_{ccc}$+$\eta_{b}$ that makes the binding.
}

For $J^{P}=3/2^{-}$, the $cccb\bar{b}$ pentaquark state
$\rm P_{c^{3}b\bar{b}}(14373,\\ 0, 3/2^{-})$ has two $ccc-b\bar{b}$ decay channels, namely $\Omega_{ccc}\Upsilon$ and $\Omega_{ccc}\eta_{b}$.
The corresponding relative partial decay widths from Table \ref{value-QQQQQ} with the assumptions of Eq. (\ref{eq:gamma})  are
\begin{equation}
\Gamma_{\Omega_{ccc}\Upsilon}:\Gamma_{\Omega_{ccc}\eta_{b}}=0.1:1.
\end{equation}
Our results suggest that $\Omega_{ccc}\eta_{b}$ is the dominant decay channel in the $ccc-b\bar{b}$ decay mode.
Moreover, the $\rm P_{c^{3}b\bar{c}}(14373, 0, 3/2^{-})$ state also has three decay channels in the $ccb-c\bar{b}$ decay mode.
Their relative partial decay widths are
\begin{equation}
\Gamma_{\Omega^{*}_{ccb}B^{*}_{c}}:\Gamma_{\Omega^{*}_{ccb}B_{c}}:\Gamma_{\Omega_{ccb}B^{*}_{c}}=3.3:3.7:1,
\end{equation}
i.e., the partial decay widths of the $\Omega^{*}_{ccb}B^{*}_{c}$ and $\Omega^{*}_{ccb}B_{c}$ channels are larger than that of the $\Omega_{ccb}B^{*}_{c}$.

The $J^{P}=1/2^{-}$ pentaquark states $\rm P_{c^{3}b\bar{b}}(14411, 0,\\ 1/2^{-})$ and $\rm P_{c^{3}b\bar{b}}(14357, 0, 1/2^{-})$ have different decay behaviors. The $\rm P_{c^{3}b\bar{b}}(14411, 0, 1/2^{-})$ can decay into the $\Omega^{*}_{ccb}B_{c}^{*}$, while this channel is kinetically forbidden for the $\rm P_{c^{3}b\bar{b}}(14357, 0, 1/2^{-})$ state.
Meanwhile, the partial decay widths for the $\rm P_{c^{3}b\bar{b}}(14411, 0, 1/2^{-})$ state has
\begin{equation}
\Gamma_{\Omega^{*}_{ccb}B_{c}^{*}}:\Gamma_{\Omega_{ccb}B^{*}_{c}}:\Gamma_{\Omega_{ccb}B_{c}}=31:4:1,
\end{equation}
and for the $\rm P_{c^{3}b\bar{b}}(14357, 0, 1/2^{-})$ state,
\begin{equation}
\Gamma_{\Omega^{*}_{ccb}B_{c}^{*}}:\Gamma_{\Omega_{ccb}B^{*}_{c}}:\Gamma_{\Omega_{ccb}B_{c}}=0:2.1:1.
\end{equation}

\subsection{The $ccbb\bar{Q}$ pentaquark states}
The $ccbb\bar{c}$ ($ccbb\bar{b}$) subsystem has eight possible rearrangement decay channels, including $\Omega^{*}_{ccb}B^{*}_{c}$ $(\Omega^{*}_{ccb}\Upsilon)$, $\Omega_{ccb}B^{*}_{c}$ $(\Omega_{ccb}\Upsilon)$, $\Omega^{*}_{ccb}B_{c}$ $(\Omega^{*}_{ccb}\eta_{b})$, $\Omega_{ccb}B_{c}$ $(\Omega_{ccb}\eta_{b})$, $\Omega^{*}_{cbb}J/\psi$ $(\Omega^{*}_{cbb}B^{*}_{c})$, $\Omega_{cbb}J/\psi$ $(\Omega_{cbb}B^{*}_{c})$, $\Omega^{*}_{cbb}\eta_{c}$ $(\Omega^{*}_{cbb}B_{c})$, and $\Omega_{cbb}\eta_{c}$ $(\Omega_{cbb}B_{c})$.

According to Fig. \ref{fig-QQQQQ} (j), the $\rm P_{c^2b^2\bar{b}}(17477,0,5/2^-)$ state does not have S-wave strong decay channels, and thus this state is expected to be narrow. It can still decay into the D-wave final states of $\Omega_{ccb}\eta_b$ and $\Omega_{cbb}B_c$.

Meanwhile, the lowest $J^P=3/2^{-}$ $ccbb\bar{b}$ pentaquark state $\rm P_{c^2b^2\bar{b}}(17416,0,3/2^-)$ is below all allowed strong decay channels. It should decay through the electromagnetic and weak interactions rather than the strong interaction. Thus, this state is considered a good stable pentaquark.

In principle, the values of $A_{i}$j and $v_{ij}$ in the modified CMI model should be different for various systems. However,
it is difficult to exactly calculate these parameters for a given system without knowing the spatial wave function. Thus, they are extracted from the masses of conventional hadrons by assuming that quark-(anti)quark interactions are the same for all the hadron systems. Of course, this assumption certainly leads to uncertainties on mass estimations for multiquark states. Since the size of a multiquark state is expected to be larger than that of a conventional hadron and the distance between quark components may be larger, the attraction between quark components should be weaker. Thus, our framework may produce a little more binding.

{To check such an effect from the difference of couplings between conventional hadron and multiquarks, we show the change of the $J^{P}=3/2^{-}$ $ccbb\bar b$ mass spectra
by varying the chromomagnetic couplings $v_{ij}$: $v_{ij}\to 0.5 v_{ij}$ or $v_{ij}\to 2 v_{ij}$. The results are shown in Table \ref{mass-ccbbb}.
From the table, the bigger the $v_{ij}$ are, the larger the mass gaps are and the smaller the lowest-lying states are. However, the variation of $v_{ij}$ from $0.5\sim2$ $v_{ij}$ would not change the conclusions in this manuscript.

\begin{table}[b]
\centering \caption{The change of the $ccbb\bar{b}$ mass spectrum by varying the chromomagnetic couplings.
}\label{mass-ccbbb}
\renewcommand\arraystretch{1.25}
\begin{tabular}{c|ccc}
\bottomrule[1.5pt]
\bottomrule[0.5pt]
\multicolumn{4}{c}{$ccbb\bar{b}$ $J^{P}=\frac{3}{2}^{-}$}\\
$v_{ij}\to$&$2v_{ij}$&$v_{ij}$&0.5$v_{ij}$\\
\bottomrule[0.5pt]
$m$ &
$\begin{pmatrix}17583\\17509\\17473\\17384\end{pmatrix}$&
$\begin{pmatrix}17554\\17479\\17457\\17416\end{pmatrix}$&
$\begin{pmatrix}17542\\17463\\17449\\17431\end{pmatrix}$\\
\bottomrule[0.5pt]
\bottomrule[1.5pt]
\end{tabular}
\end{table}
}

The lowest $J^{P}=1/2^{-}$ state, $\rm P_{c^{2}b^{2}\bar{b}}(17405, 0, 1/2^{-})$, can only decay into $\Omega_{ccb}\eta_{b}$, and its mass is slightly larger than the $\Omega_{ccb}\eta_{b}$ threshold. Thus, its width should be narrow due to the small decay phase space.

For the other three $J^P=1/2^{-}$ pentaquark states, the $\rm P_{c^{2}b^{2}\bar{b}}(17437, 0, 1/2^{-})$ state only decays into  $\Omega_{ccb}\eta_{b}$ in two-body strong decay.
The $\rm P_{c^{2}b^{2}\bar{b}}(17576, 0, 1/2^{-})$ and $\rm P_{c^{2}b^{2}\bar{b}}(17496, 0, 1/2^{-})$ states both have two different decay modes: $bbc-c\bar{b}$ and $ccb-b\bar{b}$.
They can decay freely into many allowed decay channels, and therefore they both have broad widths.
In particular, the $\rm P_{c^{2}b^{2}\bar{b}}(17576, 0, 1/2^{-})$ state has
\begin{equation}
\Gamma_{\Omega^{*}_{cbb}B^{*}_{c}}:\Gamma_{\Omega_{cbb}B^{*}_{c}}:\Gamma_{\Omega_{cbb}B_{c}}=5.6:0.5:1,
\end{equation}
and
\begin{equation}
\Gamma_{\Omega^{*}_{ccb}\Upsilon}:\Gamma_{\Omega_{ccb}\Upsilon}:\Gamma_{\Omega_{ccb}\eta_{b}}=48:7.3:1.
\end{equation}
Moreover, all the $J^{P}=1/2^{-}$ $ccbb\bar{b}$ pentaquark states can decay into $\Omega_{ccb}\eta_{b}$ final states, and this decay channel is crucial to finding $J^P=1/2^-$ $ccbb\bar{b}$ pentaquark states.

For three unstable $J^{P}=3/2^{-}$ pentaquark states, the $\rm P_{c^{2}b^{2}\bar{b}}(17457, 0, 3/2^{-})$ state has a $ccb-b\bar{b}$ decay mode and the relative decay widths are
\begin{equation}
\Gamma_{\Omega^{*}_{ccb}\eta_{b}}:\Gamma_{\Omega_{ccb}\Upsilon}=0.3:1.
\end{equation}
The other two $J^{P}=3/2^{-}$ pentaquark states both have two decay modes: $cbb-c\bar{b}$ and $ccb-b\bar{b}$.
The $\rm P_{c^2b^2\bar{b}}(17479,0,3/2^-)$ only decays into the $\Omega_{bbc}^*B_{c}$ final state in $bbc-c\bar{b}$ decay mode,
while in the $ccb-b\bar{b}$ decay mode it has
\begin{equation}
\Gamma_{\Omega^{*}_{ccb}\eta_{b}}:\Gamma_{\Omega_{ccb}\Upsilon}=0.4:1.
\end{equation}
The heaviest $J^P=3/2^-$ state, $\rm P_{c^2b^2\bar{b}}(17554,0,3/2^-)$, can easily decay into many two-body baryon-meson channels due to its large decay phase space.

\section{Summary}\label{sec8}
More and more exotic multiquark candidates are lastingly discovered in experiments these past few years  \cite{Chen:2016qju,Liu:2019zoy,Guo:2017jvc,Brambilla:2019esw}. 
The $P_c(4312)$, $P_c(4440)$, and $P_c(4457)$ states and the fully charmed tetraquark candidate $X(6900)$ reported from the LHCb Collaboration motivate us to discuss the possible pentaquark states with $QQQQ\bar{Q}$ configuration in the framework of the CMI model.


In this work, we first construct the wave functions $\psi_{flavor}\otimes\psi_{color}\otimes\psi_{spin}$ based on the SU(2) and SU(3) symmetry and the Pauli Principle.
Then, we extract the effective coupling constants from the conventional hadrons. After that, we systematically calculate the CMI Hamiltonian for the $QQQQ\bar{Q}$ pentaquark states and obtain the corresponding mass spectra in the reference system scheme.
In the modified CMI scheme, the effect of chromoelectric interaction is explicitly added.

The mass spectra is studied for the $QQQQ\bar{Q}$ pentaquark system.
In addition, we also provide the eigenvectors to extract useful information about the decay properties for the $QQQQ\bar{Q}$ pentaquark systems.
The overlaps for the pentaquark state with a particular baryon $\otimes$ meson state are obtained.
Finally, we analyze the stability, possible quark rearrangement decay channels, and relative partial decay widths for all the $QQQQ\bar{Q}$ pentaquark states.



According to our calculations and analysis, we only find two $cccc\bar{c}$ states due to the constraint from Pauli principle: a $J^{P}=3/2^{-}$ state, $\rm P_{c^{4}\bar{c}}$(7864, $0$, $3/2^{-}$), and a $J^{P}=1/2^{-}$ state $\rm P_{c^{4}\bar{c}}$(7949, $0$, $1/2^{-}$),
and there exists a no ground $J^{P}=5/2^{-}$ $cccc\bar{c}$  pentaquark state.
The same situation also happens in the $cccc\bar{b}$, $bbbb\bar{c}$, and $bbbb\bar{b}$ subsystems.
From the obtained tables and figures for the $QQQQ\bar{Q}$ pentaquark system, we find one good stable candidate: the $\rm P_{c^{2}b^{2}\bar{b}}(17416, 0, 3/2^{-})$ state.
It lies only below the allowable decay channel $\Omega_{ccb}^{*}\eta_{b}$ 4 MeV, and thus can only decay through electromagnetic or weak interactions.
Meanwhile, the $\rm P_{c^2b^2\bar{b}}(17477,0,5/2^-)$ state is also a relatively stable pentaquark since it is lower than all possible S-wave strong decay channels. It can still decay into $\Omega_{ccb}\eta_b$ and $\Omega_{cbb}B_c$ final states via the $D$-wave.

Our systematic study can provide some understanding toward these pentaquark systems. We find some fully heavy pentaquark states can be very narrow and stable. If they do exist, identifying them may not be difficult from their exotic quantum numbers and masses. The $X(6900)$ is found in the invariant mass spectrum of $J/\psi$ pairs, where two pairs of $c\bar{c}$ are produced. In our calculation, the lowest fully heavy pentaquark state is the $J^{P}=3/2^{-}$ $cccc\bar{c}$ state. To produce the lightest $cccc\bar{c}$ pentaquark state, one needs to simultaneously produce at least four pairs of $c\bar{c}$, and this seems to be a difficult task in the experiment.

{
In Ref. \cite{SilvestreBrac:1992mv}, the energies of diquonia $Q^{2}\bar{Q}^{2}$ are systemically calculated and compared to the threshold energies with orbital angular momentum $L=0$ within the framework including the chromomagnetic interactions. As pointed out there and in other references \cite{Leandri:1989su,SilvestreBrac:1992yg,SilvestreBrac:1993sb},
a more detailed study
is further needed to confirm the bound or resonant states obtained with the chromomagnetic
interaction model. Our results should similarly be checked in the future with a more serious five-body estimate in the quark model.

In conclusion, we give a preliminary prediction about the mass spectra of fully heavy pentaquarks.
More detailed dynamical investigations on the $QQQQ\bar{Q}$ pentaquark systems are still needed.
We hope that our study may inspire theorists and experimentalists to pay attention to this kind of pentaquark system.

}


\section*{ACKNOWLEDGMENTS}
This work is supported by the China National Funds for Distinguished Young Scientists under Grant No. 11825503, the National Key Research and Development Program of China under Contract No. 2020YFA0406400, the 111 Project under Grant No. B20063, and the National Natural Science Foundation of China under Grant No. 12047501. This project is also supported by the National Natural Science Foundation of China under Grants No. 11705072 and  No. 11965016, and the CAS Interdisciplinary Innovation Team.

\appendix
\section{Some expressions in detail}\label{sec10}
\newsavebox{\tablebox}
\begin{table*}[t]
\caption{The expressions of CMI Hamiltonians for $cccc\bar{c}$, $cccb\bar{c}$, and $ccbb\bar{c}$ pentaquark subsystems. The $J$ represents the spin of the pentaquark states.} \label{nnnsQ} \centering
\begin{lrbox}{\tablebox}
\renewcommand\arraystretch{1.05}
\begin{tabular}{|c|c|}
\midrule[1.5pt]
\bottomrule[0.5pt]
$J$&The expressions of CMI Hamiltonian for $cccc\bar{c}$ subsystems \\
\hline
$J=3/2$&$\frac{56}{3}C_{cc}-\frac{16}{3}C_{c\bar{c}}$\\
$J=1/2$&$\frac{56}{3}C_{cc}+\frac{32}{3}C_{c\bar{c}}$\\ \cline{1-2}
$J$&The expressions of CMI Hamiltonian for $cccb\bar{c}$ subsystems \\ \cline{1-2}
$J=5/2$&$8C_{cc}+\frac{16}{3}C_{b\bar{c}}$\\
$J=3/2$&$\begin{pmatrix}
\begin{pmatrix}\frac{28}{3}C_{cc}+\frac{28}{3}C_{cb}\\-4C_{c\bar{c}}-\frac{4}{3}C_{b\bar{c}}\end{pmatrix}&
\frac{2\sqrt{2}}{3}\begin{pmatrix}-C_{cc}+C_{cb}\\+C_{c\bar{b}}-C_{b\bar{c}}\end{pmatrix}&
-\frac{8\sqrt{5}}{3}(C_{c\bar{c}}-C_{b\bar{c}})\\

\frac{2\sqrt{2}}{3}\begin{pmatrix}-C_{cc}+C_{cb}\\+C_{c\bar{b}}-C_{b\bar{c}}\end{pmatrix}&
\begin{pmatrix}\frac{26}{3}C_{cc}-6C_{cb}\\+\frac23C_{c\bar{b}}-2C_{b\bar{c}}\end{pmatrix}&
\frac{4\sqrt{10}}{3}(C_{c\bar{c}}+2C_{b\bar{c}})\\

-\frac{8\sqrt{5}}{3}(C_{c\bar{c}}-C_{b\bar{c}})&
\frac{4\sqrt{10}}{3}(C_{c\bar{c}}+2C_{b\bar{c}})&
8(C_{cc}-C_{b\bar{c}})
\end{pmatrix}$\\
$J=1/2$&$\begin{pmatrix}
\begin{pmatrix}\frac{28}{3}C_{cc}+\frac{28}{3}C_{cb}\\+8C_{c\bar{c}}+\frac83C_{b\bar{c}}\end{pmatrix}&
\frac{2\sqrt{2}}{3}\begin{pmatrix}-C_{cc}+C_{cb}\\-2C_{c\bar{c}}+2C_{b\bar{c}}\end{pmatrix}&
\frac{2\sqrt{2}}{3}(C_{c\bar{c}}-C_{b\bar{c}})\\

\frac{2\sqrt{2}}{3}\begin{pmatrix}-C_{cc}+C_{cb}\\-2C_{c\bar{c}}+2C_{b\bar{c}}\end{pmatrix}&
\begin{pmatrix}\frac{26}{3}C_{cc}-6C_{cb}\\-\frac43C_{c\bar{c}}+4C_{b\bar{c}}\end{pmatrix}&
-\frac{2}{3}(13C_{c\bar{c}}-C_{b\bar{c}})\\

\frac{2\sqrt{2}}{3}(C_{c\bar{c}}-C_{b\bar{c}})&
-\frac{2}{3}(13C_{c\bar{c}}-C_{b\bar{c}})&
10(C_{cc}-C_{cb})
\end{pmatrix}$\\
\cline{1-2}
\Xcline{1-2}{0.7pt}
$J$&The expressions of CMI Hamiltonian for $ccbb\bar{c}$ subsystems \\ \cline{1-2}
$J=5/2$&$\frac83(C_{cc}+C_{bb}+C_{cb}+C_{c\bar{b}}+C_{b\bar{c}})$\\
$J=3/2$&$\begin{pmatrix}
\begin{pmatrix}\frac{28}{9}C_{cc}+\frac{28}{9}C_{bb}+\frac{112}{9}C_{cb}\\-\frac83C_{c\bar{c}}-\frac83C_{b\bar{c}}\end{pmatrix}&
\frac{2}{3}\sqrt{\frac23}\begin{pmatrix}C_{cc}-C_{bb}\\-2C_{c\bar{c}}+2C_{b\bar{c}}\end{pmatrix}&
-\frac{2\sqrt{2}}{9}\begin{pmatrix}C_{cc}+C_{bb}+2C_{cb}\end{pmatrix}&
\frac{16}{3}\sqrt{\frac53}(C_{c\bar{c}}-C_{b\bar{c}})\\

\frac{2}{3}\sqrt{\frac23}\begin{pmatrix}C_{cc}-C_{bb}\\-2C_{c\bar{c}}+2C_{b\bar{c}}\end{pmatrix}&
\begin{pmatrix}\frac{10}{3}C_{cc}+\frac{10}{3}C_{bb}-4C_{cb}\\-\frac23C_{c\bar{c}}-\frac23C_{b\bar{c}}\end{pmatrix}&
-\frac{2}{3\sqrt{3}}\begin{pmatrix}C_{nn}-C_{bb}+\\7C_{c\bar{c}}-7C_{b\bar{c}}\end{pmatrix}&
2\sqrt{10}(C_{c\bar{c}}+C_{b\bar{c}}) \\

-\frac{2\sqrt{2}}{9}\begin{pmatrix}C_{cc}+C_{bb}+2C_{cb}\end{pmatrix}&
-\frac{2}{3\sqrt{3}}\begin{pmatrix}C_{nn}-C_{bb}+\\7C_{c\bar{c}}-7C_{b\bar{c}}\end{pmatrix}&
\begin{pmatrix}\frac{26}{9}C_{cc}+\frac{26}{9}C_{bb}-\frac{100}{9}C_{cb}\\+\frac{10}{3}C_{c\bar{c}}+\frac{10}{3}C_{b\bar{c}}\end{pmatrix}&
-\frac{2}{3}\sqrt{\frac{10}{3}}(C_{c\bar{c}}-C_{b\bar{c}})\\

\frac{16}{3}\sqrt{\frac53}(C_{c\bar{c}}-C_{b\bar{c}}) &
2\sqrt{10}(C_{c\bar{c}}+C_{b\bar{c}})&
-\frac{2}{3}\sqrt{\frac{10}{3}}(C_{c\bar{c}}-C_{b\bar{c}})&
\begin{pmatrix}\frac83C_{cc}+\frac83C_{bb}+\frac83C_{cb}\\-4C_{c\bar{c}}-4C_{b\bar{c}}\end{pmatrix}
\end{pmatrix}$
\\
$J=1/2$&$\begin{pmatrix}
\begin{pmatrix}\frac{28}{9}C_{cc}+\frac{28}{9}C_{bb}+\frac{112}{9}C_{cb}\\+\frac{16}{3}C_{c\bar{c}}+\frac{16}{3}C_{b\bar{c}}\end{pmatrix}&
\frac{2}{3}\sqrt{\frac23}\begin{pmatrix}C_{cc}-C_{bb}\\+4C_{c\bar{c}}-4C_{b\bar{c}}\end{pmatrix}&
-\frac{2\sqrt{2}}{9}(C_{cc}+C_{bb}+2C_{cb})&
-\frac{4}{3}\sqrt{\frac23}(C_{c\bar{c}}-C_{b\bar{c}})\\

\frac{2}{3}\sqrt{\frac23}\begin{pmatrix}C_{cc}-C_{bb}\\+4C_{c\bar{c}}-4C_{b\bar{c}}\end{pmatrix}&
\begin{pmatrix}\frac{10}{3}C_{cc}+\frac{10}{3}C_{bb}-4C_{cb}\\+\frac43C_{c\bar{c}}+\frac43C_{b\bar{c}}\end{pmatrix}&
-\frac{2}{3\sqrt{3}}\begin{pmatrix}C_{cc}-C_{bb}-14\\C_{c\bar{c}}+14C_{b\bar{c}}\end{pmatrix}&
-4(C_{c\bar{c}}+C_{b\bar{c}}) \\

-\frac{2\sqrt{2}}{9}(C_{cc}+C_{bb}+2C_{cb})&
-\frac{2}{3\sqrt{3}}\begin{pmatrix}C_{cc}-C_{bb}-14\\C_{c\bar{c}}+14C_{b\bar{c}}\end{pmatrix}&
\begin{pmatrix}\frac{26}{9}C_{cc}+\frac{26}{9}C_{bb}-\frac{100}{9}C_{cb}\\-\frac{20}{3}C_{c\bar{c}}-\frac{20}{3}C_{b\bar{c}}\end{pmatrix}&
\frac{28}{3\sqrt3}(C_{c\bar{c}}-C_{b\bar{c}})\\

-\frac{4}{3}\sqrt{\frac23}(C_{c\bar{c}}-C_{b\bar{c}}) &
-4(C_{c\bar{c}}+C_{b\bar{c}})&
\frac{28}{3\sqrt3}(C_{c\bar{c}}-C_{b\bar{c}})&
\frac83(C_{cc}+C_{bb}-2C_{cb})
\end{pmatrix}$\\
\bottomrule[0.5pt]
\midrule[1.5pt]
\end{tabular}
\end{lrbox}\scalebox{0.96}{\usebox{\tablebox}}
\end{table*}


\end{document}